\documentclass[aps,pre,twocolumn,square,numbers,amssymb,amsmath,nobibnotes,footinbib]{revtex4}
\usepackage{bm}
\usepackage{amsmath}
\usepackage{amsfonts}
\usepackage{amssymb}   
\usepackage{eucal}
\usepackage{verbatim}
\usepackage{times}
\usepackage{graphicx}
\usepackage{subfigure}
\usepackage{dsfont}
\usepackage{bbold}
\usepackage{esint}
\usepackage{color}
\usepackage{enumitem}

\newcommand{\barr}{\begin{eqnarray}}
\newcommand{\earr}{\end{eqnarray}}

\newcommand{\zb}{\bar{z}}

\newcommand{\idm}{\mathbf{1}}

\usepackage{color}

\begin{document}

\title{Universal transient behavior in large dynamical systems on networks}

\author{Wojciech Tarnowski$^1$, Izaak Neri$^2$ and Pierpaolo Vivo$^2$}
\affiliation{$^1$ Marian Smoluchowski Institute of Physics, Uniwersytet Jagiello\'{n}ski, Krakow (Poland)\\
$^2$ Department of Mathematics, King's College London, Strand, London, WC2R 2LS (United Kingdom)}

\date{\today}

\begin{abstract}
We  analyze how  the transient dynamics of large dynamical systems in the vicinity of a stationary point, modeled by a set of randomly coupled linear differential equations, depends on the network topology.   We characterize the transient response  of a system  through the   evolution   in time of the squared norm of the state  vector, which is  averaged  over different realizations of the initial perturbation.   We develop a mathematical formalism that computes this quantity for graphs that are locally tree-like.   We show that for unidirectional networks the theory simplifies and general analytical results can be derived.    For example, we derive analytical expressions for the average squared norm for random directed graphs with a prescribed degree distribution.  These analytical results reveal that unidirectional systems exhibit a high degree of universality in the sense that the average squared norm only depends on a single parameter encoding the average interaction strength between the individual constituents.   In addition, we derive analytical expressions for the average squared norm for unidirectional systems with fixed diagonal disorder and with bimodal diagonal disorder.     We illustrate these results with numerical experiments on large random graphs and  on  real-world networks.
\end{abstract}


\maketitle

\section{Introduction}
Networks of interacting constituents appear in the study of systems as diverse as ecosystems \cite{Bascompte2007, Ings2009, Allesina2015}, neural networks \cite{Sompolinski1988,KuehnBook,Brunel2000, Sporns2004, Bullmore2009}, financial markets \cite{Franklin2008, Haldane2011, Bardoscia2017, Caccioli2018}, and signaling networks \cite{Jordan2000, Alon2006, Chittaranjan2019}; for more examples see  Refs.~\cite{Newman2010, Dorogov2013}.   Traditionally, a strong focus has been put on  whether such systems are stable at long time scales \cite{May1972, Allesina2012} because stability is often associated to functionality, e.g.,    stable ecosystems or economies \cite{Moran2019}.    Differently,   the short-time transient response of networked systems  to perturbations is  less understood  despite being important for applications: for example, neuroscientists administer magnetic stimulations to the brain and observe distinct dynamical responses of electrical activity, which capture different connectivity states of the underlying network of neurons \cite{Massimini2005};      in  ecological systems, the asymptotic dynamics does not capture the typical time scales accessible in experiments \cite{Neubert1997, Hastings2004, David2005, Tang2014, Arnoldi2018};     in the context of epidemics, the initial time window before vaccinations become available, makes a crucial difference in limiting the extent of the outbreak \cite{Satorras2015}.   A relevant question is thus how  network topology  determines the early-time dynamics of large systems.    For example: (i) how long does a stable system take to return to its stationary state as a function of the network topology and interaction strength among its constituents, and (ii) how long does it take to realize that a seemingly stable system is unstable after all, and disaster is looming?

We describe the state of a large dynamical system at time $t$ with  $N$ real-valued variables $\zeta_i(t)$. 
   For example, $\zeta_i(t)$ may represent the abundance of species $i$ in an ecosystem or the activity of the $i$-th neuron in the brain at time $t$. We assume that the system evolves according to a system of  first-order equations
\begin{equation}
\partial_t \zeta_i = f_i(\bm \zeta)\ ,
\end{equation}
for $i=1,\ldots,N$.
Although the functions $f_i$  are arbitrary, the dynamics can be linearized close to a stationary point $\bm\zeta^\star$   for which  $f_i(\bm\zeta^\star)=0$ to yield \cite{HG1,HG2}
\begin{equation}
\frac{\mathrm{d}y_i(t)}{\mathrm{d}t}=\sum_{k=1}^N A_{ik}y_k(t)\ ,\label{dynamical1}
\end{equation}
where the $N$-dimensional vector 
\begin{equation}
\bm y(t)=\bm\zeta(t)-\bm\zeta^\star
\end{equation}
 encodes deviations from the stationary state and 
\begin{equation}
 A_{ik}=\left(\frac{\partial f_i}{\partial \zeta_k}\right)\Big|_{\bm\zeta^\star}
\end{equation}
 are the entries of the corresponding Jacobian matrix $A$.

The magnitude of the deviation vector $\bm y(t)$ as a function of time is captured by the 
squared norm $|\bm y(t)|^2$.    In order to grasp the dynamical response to  generic initial perturbations, we consider the average 
\begin{equation}
S_{N}(t;A) = \langle |\bm y(t)|^2\rangle \label{eq:SNta}
\end{equation}     
 over initial states $\bm y(0)$   uniformly drawn from the sphere $|\bm y(0)|=\alpha$, where $\alpha$ quantifies the strength of the initial kick; without loss of generality we set $\alpha=1$.       In addition, to capture the properties of a typical dynamical system, we take $A$  to be a random matrix of pairwise interactions \cite{May1972,Allesina2015, Fyodorov2016}, and we further average the squared norm over the disorder
\begin{equation}
S_N(t)=\overline{\langle |\bm y(t)|^2\rangle}\ .  \label{eq:SNTt}
\end{equation}     
Since we will be interested in large systems, we take the limit 
  \begin{equation}
S(t)=\lim_{N\to\infty}S_N(t)\ .  
\end{equation}  
If  
\begin{equation}
\lim_{t\rightarrow \infty}S(t) = 0 \quad (\lim_{t\rightarrow \infty}S(t) = \infty)\ ,
\end{equation}
then we say that a system is {\it transiently (un)stable}.  Note that  transient stability is different from   {\it asymptotic stability}, which is governed by the sign of the real part of the leading eigenvalue  of~$A$, see e.g. Refs.~\cite{May1972,Allesina2015,Neri2019}.    

The observables $S_N(t)$ and $S(t)$ are  the main objects of interest.     
The quantity $S(t)$ describes the transient dynamics of an infinitely large dynamical system and can be computed analytically for a large and important class of systems, as we show in this paper.  On the other hand,  $S_N(t)$ captures the dynamics of systems at finite $N$ and  is the quantity that is experimentally measurable.    There exists a crossover time $t^\star(N)$ such that for 
 times $t<t^\star(N)$ theory and experiment are in correspondence, i.e.~$S_N(t)\approx S(t)$, while  for $t\gg t^\star(N)$ theory and experiment are in disagreement, i.e.~$S_N(t)\gg S(t)$. The  discrepancy between $S(t)$ and $S_N(t)$ at large $t$ is most evident for transiently stable systems, for which $\lim_{t\rightarrow \infty}S_N(t) = \infty$ but $\lim_{t\rightarrow \infty}S(t) = 0$.
 
 For transiently stable systems, where $S_N(t)$ has a non-monotonic behavior, we define the crossover time by
\begin{equation} 
t^\star(N) = {\rm min}_{t\geq 0} S_N(t)\ .\label{tstardef}
\end{equation}
The crossover time $t^\star(N)$  defined by Eq. \eqref{tstardef} increases with $N$ and diverges for large $N$ [seemingly as a power law, see Appendix \ref{App:NdependN} for a detailed discussion]. For the reasons above, $S(t)$ is a good measure of the transient dynamics of   large dynamical systems.

So far, the quantity $S(t)$  has only been  computed    for systems with \emph{fully connected} topology \cite{Mehlig2000,Grela2017,Erdos2018,Marti2018, Erdos2019}, namely for 
\begin{equation}
A_{ij}=-\mu \delta_{ij}+X_{ij}/\sqrt{N}\ ,
\end{equation}
where the  $X_{ij}$  are independent and identically distributed (i.i.d.)~entries with zero mean and finite moments.     In this case, 
\begin{equation}
S(t)=e^{-2\mu t}I_0(2\rho t) \stackrel{t\rightarrow \infty}{\sim}  \frac{e^{2 t(\rho-\mu)}}{\sqrt{4\pi \rho t}} \ ,\label{Sfullyconnected}
\end{equation}
where $\rho$ is the spectral radius of the matrix $X/\sqrt{N}$ and $I_0(x)$ is the modified Bessel function of the first kind.      
Since $S(t)$ only depends on the spectral radius of the matrix $X$, it enjoys a high degree of universality.

We aim to study how $S(t)$ depends on the topology of a network, for which it is  necessary to go beyond the fully connected paradigm.    We  assume  that $A$ is the adjacency matrix of a weighted graph that is locally tree-like, 
     \begin{equation}
A_{ij} = D_i \delta_{ij}+C_{ij}J_{ij}\ ,  \label{eq:modelDef}
\end{equation}  
where $D_i\in \mathbb{R}^-$  are the diagonal decay rates, the $J_{ij}\in \mathbb{R}$   are the coupling strengths,  and the $C_{ij}\in\left\{0,1\right\}$ are the entries of the adjacency matrix of a {\it locally tree-like}, directed and simple graph.      We say that a sequence of graphs is locally tree-like if in the limit $N\rightarrow \infty$ the finite  neighborhood of a randomly selected node is with probability one a tree~\cite{Dembo2010,Dembo2010x};  loosely speaking, a graph is locally tree-like if it is large and does not contain small cycles.      
In this paper, we  develop a mathematical method  to compute $S(t)$ for the  model given by Eq. \eqref{eq:modelDef} under the sole assumption that the graph represented by $C$  is locally tree-like.

We further illustrate the general mathematical formalism on a 
canonical class of random directed graphs, namely, the ensemble of adjacency matrices of weighted random graphs with a prescribed degree distribution \cite{Dorogovtsev2001, Newman2001, Dorogov2013, Metz2018, Neri2019}.   These random graphs are  used to model real-world systems, such as, the Internet \cite{Broder, Dorogovtsev2001, Pastor},  neural networks \cite{Brunel} and other high-dimensional systems \cite{Newman2001, Dorogov2013, Newman2010}.  
In this ensemble, the matrix $A$ in Eq. \eqref{eq:modelDef} is defined as follows: 
\begin{itemize}
\item the $D_i$ are i.i.d.~taken from a probability density $p_D(x)$;
\item    the $J_{ij}$ are  i.i.d.~random variables with probability density $p_J(x)$;
\item  the $C_{ij}$ are the entries of the adjacency matrix of a random (directed) graph with a  prescribed joint degree distribution  $p_{\mathrm{deg}}(k_{\mathrm{in}},k_\mathrm{out})$ of in-degrees $k_{\mathrm{in}}$ and out-degrees $k_{\mathrm{out}}$.      In addition, we assume that the degree distribution factorizes as
  \begin{equation}
p_{\mathrm{deg}}(k_{\mathrm{in}},k_\mathrm{out})=p_{\mathrm{deg}}(k_{\mathrm{in}})p_{\mathrm{deg}}(k_\mathrm{out})\ . \label{eq:prescr}
  \end{equation}

\end{itemize}  
The spectral properties of this ensemble have been studied in detail in Refs.~\cite{Rogers2009, Neri2016, Metz2018, Neri2019} and the asymptotic stability of dynamical systems described by this ensemble has been studied in   Ref.~\cite{Neri2019}.
Here, we   characterize the transient response to a random perturbation for dynamical systems on random graphs with a prescribed degree distribution by deriving analytical expressions for $S(t)$. 

This analytical progress is made possible because (i) our general theory simplifies for locally \emph{oriented} (or \emph{unidirectional}) networks (see Section \ref{orientedtrees} for a precise statement), and (ii) directed random graphs with a prescribed degree distribution are -- with high probability -- locally oriented \cite{Dembo2010,Dembo2010x}. 

Remarkably, we find that $S(t)$ is universal for directed random graphs with a prescribed degree distribution, in the sense that it only depends on the distribution $p_J(x)$  and the degree distribution $p_{\mathrm{deg}}(k)$ through a single parameter (see our main formula \eqref{finalformula}).       On the other hand, the dependence on $p_D$ is non-universal.   

We compare the derived analytical results for $S(t)$ with numerical results for $S_N(t)$ on random graphs and on real-world graphs.   We find that $S(t)$ is in excellent agreement with  $S_N(t)$ as long as $t<t^\star(N)$, where $t^\star(N)$ is a timescale that diverges with $N$.   For  $t\gg t^\star(N)$, it holds that $S_{N}(t;A)\sim e^{2 {\rm Re}[\lambda_1(A)]t}$, with $\lambda_1(A)$  the eigenvalue of $A$ with the largest real part, and     as a consequence, $S(t) \ll  S_N(t)$.

The plan of the paper is as follows. In  Sec.~\ref{sec:2point}, we express $S(t)$ as a contour integral over the two-point correlator of a random matrix ensemble, effectively mapping a dynamical systems problem onto a random matrix theory problem.     In  Sec.~\ref{sec:derivation}, we develop a mathematical formalism   to compute  the two-point correlator, and thus also $S(t)$, on   tree or locally tree-like graphs.   In Sec.~\ref{sec:oriented},   we consider directed random graphs with a prescribed  joint degree distribution and we derive for this ensemble analytical expressions for   $S(t)$.        In Sec.~\ref{sec:Empirical}, we compare the obtained analytical expressions for infinitely large graphs with numerical experiments on random graphs of finite size and on real-world graphs.   In Sec.~\ref{Disc}, we discuss the obtained results and present an outlook for future research.    In Appendix~\ref{App:NdependN}, we analyze how $t^\star(N)$ depends on $N$.    In Appendix~\ref{appB}, we compute a contour integral that appears in our calculations.    In Appendix~\ref{bimodalSpec}, we make a study of the spectra of random graphs with a prescribed degree distribution and a bimodal distribution of diagonal matrix entries, and in Appendix~\ref{powerlaw}, we  present numerical results for random graphs with power-law degree distributions.  

\section{Mapping onto a random matrix problem}\label{sec:2point} 
In this section, we derive the formula  
\begin{equation}
S_{N}(t;A) =\varoiint_\gamma \frac{\mathrm{d}z\mathrm{d}w~e^{t(z+w)}}{N(2\pi\mathrm{i})^2}\mathcal{W}_A(z,w)\ ,\label{St}
\end{equation}
which expresses the dynamical response  $S_{N}(t;A)$  as  a contour integral  of the  two-point correlator 
\begin{equation}
\mathcal{W}_A(z,w)=\mathrm{Tr}\left[\frac{1}{z\mathbb{1}-A^T}\frac{1}{w\mathbb{1}-A}\right] \label{eq:twopoint}
\end{equation} 
    over a closed counterclockwise-oriented contour $\gamma$ that encloses all eigenvalues of $A$.  The  symbol $\mathbb{1}$ denotes the identity matrix of size $N$.    
    
 Analogously, we obtain 
\begin{equation}
S(t) = \varoiint_\gamma \frac{\mathrm{d}z\mathrm{d}w~e^{t(z+w)}}{(2\pi\mathrm{i})^2}\overline{\mathcal{W}}(z,w)\ ,\label{StAverage}
\end{equation}
where
    \begin{equation}
    \overline{\mathcal{W}}(z,w)=\lim_{N\rightarrow \infty}\frac{1}{N}\overline{\mathrm{Tr}\left[(z\mathbb{1}-A^T)^{-1}(w\mathbb{1}-A)^{-1}\right]} \label{eq:av2point}
    \end{equation} 
is the average two-point correlator.  

  Formulae \eqref{St}-\eqref{eq:av2point} provide a recipe to compute $S(t)$: if we obtain an expression  
   for the  two-point correlator  \eqref{eq:av2point} of a random matrix ensemble,  then  $S(t)$ follows readily from evaluating  the contour integral \eqref{St}.       Note that we only need to compute $\overline{\mathcal{W}}(z,w)$  for values $z,w$ that lie outside the continuous part of the spectrum.
    
   In order to obtain \eqref{St}, we first  express the solution of  Eq.~\eqref{dynamical1} as
\begin{equation}
{\bm y(t)} = e^{At}{\bm y}(0)\ , 
\end{equation}
and therefore
\begin{equation}
|\bm{y}(t)|^2=\bm{y}^T(0)e^{A^Tt}e^{At}\bm{y}(0)\ .\label{S2}
\end{equation} 

Since $|\bm{y}(0)|^2=1$, there exists a matrix $O$ in the the orthogonal group $\mathrm{O}(N)$ (the group of isometries of the $N$-sphere) such that 
\begin{equation}
\bm{y}(0)=O\bm{e}_1\ ,
\end{equation} 
where 
\begin{equation}
\bm{e}_1 = (1 , 0,0,\ldots , 0)^T\ .
\end{equation}
   Taking the average of  \eqref{S2} with respect to  all initial conditions $\bm{y}(0)$ selected uniformly at random on the unit sphere is thus 
    equivalent to taking the average of the following expression
\begin{equation}
\bm{e}_1^T O^T e^{A^Tt}e^{At}O\bm{e}_1 \label{eq:oo}
\end{equation}
with respect to the uniform (Haar) measure on the orthogonal group. Using that~\cite{Collins2003} 
\begin{equation}
\left< O_{ij} O_{kl}\right>=\frac{1}{N}\delta_{ik}\delta_{jl}\ ,
\end{equation}
and Eqs. \eqref{eq:oo} and  \eqref{eq:SNta}, we obtain 
\begin{equation}
S_N(t;A)=\frac{1}{N} {\rm Tr}\, e^{A^Tt}e^{At}\ . \label{eq:yT}
\end{equation} 

We use  the Dunford-Taylor formula (see \cite{Kato}, Eq. (5.47) on page 44) to express the right hand side of Eq. \eqref{eq:yT} as a contour integral.    Let $f$ be an analytic function  on the complex plane.    The Dunford-Taylor formula states that
\begin{equation}
 f(A) = \frac{1}{(2\pi {\rm i})} \oint_{\gamma} \frac{f(z)}{z-A}\mathrm{d}z \ ,   \label{eq:Dunford}
\end{equation}   
where $\gamma$ is a closed counterclockwise-oriented contour that encompasses a region of the complex plane that contains all eigenvalues of $A$.      Using \eqref{eq:Dunford}  in \eqref{eq:yT},  we readily obtain Eqs. \eqref{St} and \eqref{eq:twopoint}.

The quantity $S_{N}(t;A)$ is determined by both the statistics of  eigenvalues  and eigenvectors of $A$.     This is most clearly seen by considering the 
 special case where $A$  is diagonalizable.  In this case, $A$ can be written in its canonical form
\begin{equation}
A =\sum_{j=1}^N \lambda_j|r_j\rangle\langle \ell_j|\ ,  
\end{equation} 
where  $\lambda_j$ are its eigenvalues, and   $|r_j\rangle$ and $\langle \ell_j|$ form a bi-orthonormal set of right and left eigenvectors.
Plugging this canonical form of $A$   in Eq. \eqref{eq:yT} we obtain
\begin{equation}
S_{N}(t;A) = \frac{1}{N}\sum_{j,k} \tilde{O}_{jk}e^{t(\lambda_j^\star+\lambda_k)}\ ,
\end{equation} 
where 
\begin{equation}
\tilde{O}_{jk}=\langle\ell_k|\ell_j\rangle\langle r_j|r_k\rangle
\end{equation}
 encode the eigenvectors overlaps \cite{Mehlig2000}.   Additionally, if  $A$ is a normal matrix ($[A,A^T]=0$),  then  $\langle r_j | r_k \rangle = \langle l_k|l_j\rangle = \delta_{jk}$, and  
\begin{equation}
\langle |\bm y(t)|^2\rangle= \frac{1}{N}\sum^{N}_{j=1} e^{2t\:{\rm Re}[\lambda_j]}\ .
 \label{StSupp2x}
\end{equation}    Therefore, the non-orthogonality of eigenvectors is a primary source of transient behavior, since the eigenmodes can interfere constructively to deliver an initial amplification of the signal well before it eventually dies out \cite{Murphy2009, Ridolfi2011, Hennequin2014, Caravelli2016, Gudowska2018,savin1997,trefethen2005,asllani2018a,asllani2018}.

Eq. \eqref{St} reduces the computation of $S(t)$ to a computation of the 
     average two-point correlator  $\overline{W}(z,w)$.   Although the  one-point correlator 
\begin{equation}
\mathcal{W}_A(z)=\mathrm{Tr}\frac{1}{z\mathbb{1}-A^T}
\end{equation} 
of sparse random graphs has been studied extensively in Ref.~\cite{Rogers2009, Neri2016, Metz2018, Neri2019}, to our knowledge, the two-point correlator has not been considered before.     In the next section, we show how    to compute  the two-point correlator  $\mathcal{W}_A$ for  tree  matrices, and in the subsequent section we compute $\overline{\mathcal{W}}$ for the canonical model of random graphs with a prescribed degree distribution.

\section{Tree graphs}\label{sec:derivation} 
In this section, we present a mathematical method to compute the  two-point correlator $\mathcal{W}_A(z,w)$,  and thus also $S_N(t;A)$,  under the sole assumption that the  graph represented by $C_{ij}$ is  a tree.    

 The mathematical method we employ is based on two ideas: the  {\it size-doubling trick}  presented in the first subsection, and a recursive implementation of the \emph{Schur formula},  presented in the second subsection.     In a third subsection we discuss how the mathematical formalism simplifies for  \emph{oriented} graphs.  We say that a graph is oriented if 
\begin{equation}
C_{ij}C_{ji} = 0
\end{equation}
for all pairs $(i,j)$.

\subsection{Size doubling trick}
We use the following identity 
\begin{equation}
\mathcal{W}_A(z,w) = -\mathrm{bTr}_{11} B^{-1}\ , \label{eq:blocktr}
\end{equation}
which expresses the trace of
\begin{equation}
\frac{1}{z\mathbb{1}-A^T}\frac{1}{w\mathbb{1}-A}
\end{equation}
in terms of the block trace of the inverse of the matrix 
\begin{equation}
B =\begin{pmatrix}
\mathbb{0} & w\mathbb{1}-A\\
z\mathbb{1}-A^T & \mathbb{1}
\end{pmatrix}\ .\label{btr}
\end{equation} 
The block trace $\mathrm{bTr}_{11}$ of a $2N\times 2N$ matrix $X$  is defined  by
\begin{equation}
\mathrm{bTr}_{11} X = \sum_{j=1}^N [X]_{j,j}\ .
\end{equation}
Since $B$ is a matrix of size $2N\times 2N$ and $A$ is a matrix of size $N\times N$, we call this the size-doubling trick, which bears some similarity with the Hermitization method \cite{Feinx, Fein} in non-Hermitian random matrix theory.

In order to proceed, we note that the block trace formula in \eqref{eq:blocktr} is similar to  the block trace formula for the  spectral density of a sparse graph, see Eq.~(57) in Ref.~\cite{Metz2018}.   In the next subsection we will exploit this mathematical similarity.

\subsection{Recursion relations}       
We derive a set of recursion relations that will provide us with the diagonal elements $[B^{-1}]_{j,j}$.    The recursions can be closed using  that $A$ is the adjacency matrix of a (weighted) tree.  In particular, we use the following property.  Let $A^{(j)}$ be the matrix obtained from  $A$ by removing the $j$-th row and column; $A^{(j)}$  is the adjacency matrix of the so-called  cavity graph obtained by removing the $j$-th node from the original graph \cite{Mezard2001, Mezard2003, Rogers2008}.   It then holds that $A^{(j)}$ is a forest of $|\partial_j|$ isolated trees, where we have used the notation 
\begin{equation}
\partial_j = \left\{i: C_{ij} \neq 0 \quad {\rm or} \quad C_{ji}\neq 0\right\}
\end{equation}
for the neighborhood of $j$ and $|\partial_j|$ is the number of elements in $\partial_j$.  We use the  Schur inversion formula to derive the recursion relations.  However, first we  need to define the quantities that appear in the recursion, which requires a reshuffling of the elements of $B$.

\subsubsection{Preliminary ordering of matrix elements}

 The matrix $B$ can be seen as a replicated version of the original matrix $A$, where each node of the original graph has two replicas, one with label $i$ and the other with label $i+N$ ($i=1,\ldots,N$), which are located therefore far apart in the matrix $B$.   
Therefore,  we operate on the $2N\times 2N$ matrix $B$ to create a new matrix that preserves the same connectivity structure of the original graph (encoded in $C$). This is achieved by bundling together labels that refer to the same node.

More precisely, we  permute the rows and columns of the matrix $B$. The permutation we perform defines the matrix $\tilde B$, whose 
entries are assigned according to the following operations: 
\begin{equation}
B_{i,j}\to
\begin{cases}
\tilde{B}_{2i-1,2j-1}, &\text{ if }1\leq i,j\leq N,\\
\tilde{B}_{2(i-N),2j-1}, &\text{ if }N+1\leq i\leq 2N,1\leq j\leq N,\\
\tilde{B}_{2i-1,2(j-N)}, &\text{ if }1\leq i\leq N,N+1\leq j\leq 2N,\\
\tilde{B}_{2(i-N),2(j-N)}, &\text{ if }N+1\leq i,j\leq 2N\ .
\end{cases}
\end{equation}
This permutation is performed through a similarity transformation $\tilde{B}=P^{T}BP$, where $P$ is a suitably defined permutation matrix.

We then obtain the permuted matrix $\tilde{B}$ which consists of diagonal $2\times 2$ blocks (labelled using the sans-serif font)
\begin{equation}
\tilde{\mathsf{B}}_{ii} =  \begin{pmatrix}  0 &  w-A_{ii} \\ z -A_{ii} & 1  \end{pmatrix} :=\begin{pmatrix}  0 &  w \\ z & 1  \end{pmatrix}-\mathsf{A}_{ii}\ ,\label{Bsftilde}
\end{equation}
and  off-diagonal blocks of the form 
\begin{equation}
\tilde{\mathsf{B}}_{ij}  = \begin{pmatrix}  0 &  -A_{ij} \\ -A_{ji}& 0  \end{pmatrix} :=-\mathsf{A}_{ij}\ ,\label{defJ}
\end{equation}
for $i,j = 1,\ldots,N$. Note that now the matrix $\tilde B$ is a block matrix (formed by $2\times 2$ blocks) that inherits the same connectivity structure as the matrix $A$ (or $C$), since elements of $A$ referring to the same node - which were located far apart in the matrix $B$ - are now bundled together.

The elements of $B^{-1}$, which we need in Eq.~\eqref{eq:blocktr}, are related to the elements of $\tilde{B}^{-1}$ in the following way:
\begin{equation}
\tilde{B}^{-1}= (P^TBP)^{-1}\Rightarrow P\tilde{B}^{-1}P^T = B^{-1}\ ,\label{BtildeB}
\end{equation}
since a permutation matrix is an orthogonal transformation. 

Moreover, the block trace needed in Eq. \eqref{eq:blocktr} reads
\begin{align}
&\nonumber \mathrm{bTr}_{11}B^{-1} = \mathrm{bTr}_{11}[P\tilde{B}^{-1}P^T] =\\
&=\sum_{j=1}^N [P\tilde{B}^{-1}P^T] _{j,j}  =\sum_{j=1}^N [\tilde{B}^{-1}]_{2j-1,2j-1}\ ,\label{chain}
\end{align}
where in the first line we used Eq. \eqref{BtildeB}, and in the second line the fact that the permutation matrix $P$ maps indices $j$ onto $2j-1$ if $1\leq j\leq N$.

Therefore, the objects of interest are now the elements $[\tilde{B}^{-1}]_{2j-1,2j-1}$ of the inverse matrix of $\tilde B$.

Defining the $2\times 2$ matrices
$ \mathsf{G}_j$ for $j=1,\ldots, N$ as
\begin{equation}
 \mathsf{G}_j =  \begin{pmatrix} [\tilde{B}^{-1}]_{2j-1,2j-1} & [\tilde{B}^{-1}]_{2j-1,2j} \\ [\tilde{B}^{-1}]_{2j,2j-1} & [\tilde{B}^{-1}]_{2j,2j} \end{pmatrix}\ ,\label{Gj}
\end{equation}  
the  two-point correlator reads (see Eq.~\eqref{chain})
\begin{equation}
\mathcal{W}_A(z,w)=-\mathrm{bTr}_{11} B^{-1}=-\sum_{j=1}^N [ \mathsf{G}_j ]_{1,1}\ ,\label{TrG}
\end{equation}
and the  one-point resolvent  reads
\begin{equation}
\frac{1}{N}{\rm Tr}\left[\frac{1}{z\mathbb{1}-A^T}\right]    =   \frac{1}{N}\sum^{N}_{j=1}[\mathsf{G}_j ]_{1,2}
\end{equation} 
or 
\begin{equation}
 \frac{1}{N}{\rm Tr}\left[\frac{1}{w\mathbb{1}-A}\right]   =    \frac{1}{N}\sum^{N}_{j=1}[\mathsf{G}_j ]_{2,1}\ .
\end{equation} 

In the next subsection we derive a set of recursion relations for $ \mathsf{G}_j$'s using Schur inversion formula. 

\subsubsection{Schur formula }
We employ the  \emph{Schur inversion formula}
 \begin{align}
 \left(\begin{array}{cc} A& B \\ C & D \end{array}\right)^{-1} =  \left(\begin{array}{cc} S_D & -S_D B D^{-1}\\-D^{-1}CS_D & S_A\end{array}\right), \label{SchurInversion}
 \end{align}
 with $S_D=(A-BD^{-1}C)^{-1}$ the inverse of the Schur complement of $D$ and $S_A=(D-CA^{-1}B)^{-1}$ the inverse of the Schur complement of $A$.    

First, we show  how this works for $j=1$ and later we generalize for arbitrary $j$.   
In order to implement the Schur inversion formula, we represent $\tilde B$ with the block matrix structure of the form
\begin{equation}
\tilde{B} = \begin{pmatrix}
\tilde{\mathsf{B}}_{11} & \tilde{B}_{1\star}\\
\tilde{B}_{\star 1} & \tilde B^{(1)}\\
\end{pmatrix}\ ,\label{Btildeblock}
\end{equation}
where $\tilde{\mathsf{B}}_{11}$ is the $2\times 2$ matrix defined in Eq. \eqref{Bsftilde}, $\tilde{B}_{1\star}$ and $\tilde{B}_{\star 1}$ are $2\times 2(N-1)$ and $2(N-1)\times 2$ matrices respectively, and $\tilde B^{(1)}$ is a $2(N-1)\times 2(N-1)$ matrix.  
The matrix  $\tilde B^{(1)}$ is the $\tilde B$ matrix 
of $A^{(1)}$ obtained by removing the first column and row from the matrix $A$.

Now, we are ready to find an equation for $\mathsf{G}_1$, the upper-left $2\times 2$ block of $\tilde B^{-1}$ (see Eq. \eqref{Gj}), taking full advantage of the block structure in Eq.~\eqref{Btildeblock}. The Schur  formula  applied to the upper-left $2\times 2$ block of $\tilde B$ gives the following  
\begin{equation}
\mathsf{G}_1 = \frac{1}{\tilde{\mathsf{B}}_{11} -  \tilde{B}_{1\star} [\tilde B^{(1)}]^{-1} \tilde{B}_{\star 1}}\ .
\end{equation}
Now, using Eq. \eqref{Bsftilde} and the fact that both $\tilde{B}_{1\star} $ and $\tilde{B}_{\star 1} $ are concatenations of matrices of the form $\mathsf{A}$ (see Eq. \eqref{defJ}), we obtain

\begin{equation}
\mathsf{G}_1 = \frac{1}{ \begin{pmatrix} 0 & w \\ z & 1  \end{pmatrix} - \mathsf{A}_{11}   - \sum_{k\in  \partial_1}\sum_{\ell\in\partial_1} \mathsf{A}_{1k} \mathsf{G}^{(1)}_{k\ell} \mathsf{A}_{\ell 1}   }\ ,\label{Gjintermediate}
\end{equation}  
where
\begin{equation}
 \mathsf{G}_{k\ell}^{(1)} =  \begin{pmatrix} [(\tilde{B}^{(1)})^{-1}]_{2k-1,2\ell-1} & [(\tilde{B}^{(1)})^{-1}]_{2k-1,2\ell} \\ [(\tilde{B}^{(1)})^{-1}]_{2k,2\ell-1} & [(\tilde{B}^{(1)})^{-1}]_{2k,2\ell} \end{pmatrix}\ .\label{Gjk}
\end{equation} 
Note that  $ \mathsf{G}_{kk}^{(1)} =  \mathsf{G}_{k}^{(1)}$, the matrices we defined before in \eqref{Gj}.
In the sums in Eq. \eqref{Gjintermediate}, we omit contributions from $k,\ell\notin\partial_1$ because $\mathsf{A}_{1k}$ and $\mathsf{A}_{\ell 1}$ are null matrices in this case (see Eq. \eqref{defJ}).

Since $A$ is the adjacency matrix of a (weighted) tree graph, it holds that  $ \mathsf{G}_{k\ell}^{(1)}$ is null for any $k,\ell\in\partial_1$ with $k\neq \ell$.  To show this, note the following facts:
\begin{enumerate}
\item the nodes $k$ and $\ell$ belong to distinct trees in the forest represented by $A^{(1)}$;
\item the matrix $(\tilde{B}^{(1)})^{-1}$ has the mathematical form of  a resolvent matrix $(u\mathbb{1}  -  X)^{-1}$, where $X$ is a block matrix  built out of $2\times 2$ matrices located at the edges of the graph represented by $A$;
\item the matrix element $[X^{n}]_{k,\ell}$ denotes the sum of the weights of the  paths in the graph of length $n$ that connect node $k$ to node $\ell$.  If there exists no path that runs form  $k$ to $\ell$, then  $[X^{n}]_{k,\ell} = 0$ for all $n$.
\end{enumerate}
Therefore, expanding $(u\mathbb{1}  -  X)^{-1} = \sum^{\infty}_{n=0} X^{n}/u^{n+1}$, it follows that $[(u\mathbb{1}  -  X)^{-1} ]_{k,\ell} = 0$ (and therefore also $ \mathsf{G}_{k\ell}^{(1)}$) if 
$k$ and $\ell$ belong to distinct trees. 

Hence, applying that  for tree matrices the $ \mathsf{G}_{k\ell}^{(1)}$ are null if $k,\ell\in\partial_1$ with $k\neq \ell$, we can simplify 
Eq. \eqref{Gjintermediate} to
\begin{equation}
\mathsf{G}_1 = \frac{1}{ \begin{pmatrix} 0 & w \\ z & 1  \end{pmatrix} - \mathsf{A}_{11}   - \sum_{k\in  \partial_1}\mathsf{A}_{1k} \mathsf{G}^{(1)}_{k} \mathsf{A}_{k 1}   }\ .\label{Gjintermediate2}
\end{equation}  
This equation can be generalized to an arbitrary node $j$ because the Schur formula works the same way upon a permutation that is equivalent to relabelling of nodes. Hence, for an arbitrary node $j=1,\ldots,N$, Eq. \eqref{Gjintermediate2} becomes
\begin{equation}
\mathsf{G}_j = \frac{1}{ \begin{pmatrix} 0 & w \\ z & 1  \end{pmatrix} - \mathsf{A}_{jj}   - \sum_{k\in  \partial_j}\mathsf{A}_{jk} \mathsf{G}^{(j)}_{k} \mathsf{A}_{k j}   }\ . \label{Gjintermediate3}
\end{equation}  
This set of equations is not closed, because in general $\mathsf{G}_j \neq \mathsf{G}_j^{(k)}$ for $k\in\partial_j$. To close the equations, we can repeat the same procedure on the $N$  graphs $A^{(j)}$ obtained by removing one node at a time. We then obtain
\begin{equation}
\mathsf{G}^{(j)}_k = \frac{1}{ \begin{pmatrix} 0 & w \\ z & 1  \end{pmatrix} - \mathsf{A}_{kk}  - \sum_{\ell \in  \partial_k\setminus j} \mathsf{A}_{k\ell } \mathsf{G}^{(k,j)}_\ell \mathsf{A}_{\ell k}   }\ ,
\end{equation}  
where  $j\in\left\{1,2,\ldots,N\right\}, k\in \partial_j$, and $\mathsf{G}^{(k,j)}_\ell $ are defined analogously to Eq. \eqref{Gjk} but on the graph where the two nodes $k,j$ have been removed.

Note that in general  $\mathsf{G}^{(k,j)}_\ell \neq \mathsf{G}^{(k)}_\ell$, and the recursion has to be continued.  However, for tree matrices the recursion can be closed at the second step.   Indeed, since nodes $\ell$ and $j$ belong to distinct trees on the  graph $A^{(k)}$, the further removal of node $j$ does not affect $\mathsf{G}^{(k)}_\ell $ and therefore $\mathsf{G}^{(k,j)}_\ell =\mathsf{G}^{(k)}_\ell $. We finally obtain a closed set of equations 
\begin{equation}
\mathsf{G}^{(j)}_k = \frac{1}{ \begin{pmatrix} 0 & w \\ z & 1  \end{pmatrix} - \mathsf{A}_{kk}  - \sum_{\ell \in  \partial_k\setminus j} \mathsf{A}_{k\ell } \mathsf{G}^{(k)}_\ell \mathsf{A}_{\ell k}   }\ ,
\label{recfin}\end{equation}  
which are valid 
for  all $k\in\partial_j$ on tree graphs.        

Solving the Eqs.~\eqref{Gjintermediate3} and \eqref{recfin} on one  instance of a tree graph, one would get access to the corresponding two-point correlator
$
\mathcal{W}_A(z,w)=-\sum_{j=1}^N [\mathsf{G}_j]_{1,1}$. Since the Eqs.~\eqref{Gjintermediate3} and \eqref{recfin} are local equations, i.e. the right-hand sides of the  Eqs.~\eqref{Gjintermediate3} and \eqref{recfin} only depend on the local neighborhood of node $j$, we can also apply  Eqs. \eqref{Gjintermediate3} and \eqref{recfin}  to graphs that are locally tree-like.

 \subsection{Exact solution for oriented  trees}\label{orientedtrees}
 
 We show now how Eqs.~\eqref{Gjintermediate3} and \eqref{recfin} simplify in the case of oriented trees, i.e., trees for which all edges are unidirectional ($C_{jk}C_{kj} = 0$ for all $k\neq j$).    
 
In this case, the last terms in the denominator of the r.h.s. of Eqs.~\eqref{Gjintermediate3} and \eqref{recfin} have off-diagonal entries equal to zero. Therefore,
\begin{equation}
\mathsf{G}^{(j)}_k = \begin{pmatrix}  \alpha^{(j)}_k&  \frac{1}{z-D_{k}}\\ \frac{1}{w-D_{k}} & 0\end{pmatrix}\qquad
\mathsf{G}_k = \begin{pmatrix}  \alpha_k&  \frac{1}{z-D_{k}}\\ \frac{1}{w-D_{k}} & 0\end{pmatrix}\ , \label{2matricesG}
\end{equation}
where 
\begin{equation}
 \alpha^{(j)}_k =  \frac{\sum_{\ell\in\partial_k\setminus j}C_{\ell k}~\alpha^{(k)}_{\ell} J_{\ell k}^2 -1}{(z-D_{k})(w-D_{k})}\ ,\label{alphajk}
\end{equation}
and
\begin{equation}
 \alpha_k =  \frac{\sum_{\ell\in\partial_k}C_{\ell k}~\alpha^{(k)}_{\ell} J_{\ell k}^2 -1}{(z-D_{k})(w-D_{k})}\ .\label{alphak}
\end{equation}

    The recursion  relations given by  Eqs. \eqref{alphajk}-\eqref{alphak}  can be solved numerically on a fixed instance and the two-point correlator is given by 
     \begin{equation}
   \mathcal{W}_A(z,w)=- \sum^N_{k=1} \alpha_k\ ,
    \end{equation} 
whereas for  oriented trees  the  one-point resolvent reads \cite{Neri2016, Neri2019}
      \begin{equation}
   \mathcal{W}_A(z)= \sum^N_{k=1}\frac{1}{z-D_k}\ .
    \end{equation}

   In practice, we will often be interested in the dynamics on graphs that are locally tree-like and locally oriented.   Eqs. \eqref{alphajk}-\eqref{alphak}  also apply to these graph ensembles since the  right-hand side of the  Eqs. \eqref{alphajk}-\eqref{alphak} only depends on the local neighborhood of node $k$.   In the next section, we consider an important class of random graphs that are locally tree-like and locally oriented.  
     
\section{Directed random graphs with a prescribed degree distribution} \label{sec:oriented}
We derive analytical expressions for $S(t)$ for the  case when $A$ is the adjacency matrix of a directed random graph with a prescribed joint degree distribution, as defined in the introduction.      This ensemble of random graphs is locally tree-like and locally oriented, and therefore the Eqs. \eqref{alphajk}-\eqref{alphak} apply.   To summarize,  we will obtain  the general formula
 \begin{equation}
S(t) =\frac{1}{(2\pi\mathrm{i})^2}\varoiint_\gamma \frac{e^{t(z+w)}\mathrm{d}z\mathrm{d}w}{\left[\int \frac{p_D(x)\mathrm{d}x}{(z-x)(w-x)}\right]^{-1}-r^2}\ ,\label{finalformula}
 \end{equation}
where 
\begin{equation}
r^2=c\overline{J^2} \label{eq:rDef}
\end{equation} 
 is the product of the mean out-degree (or in-degree)  
\begin{equation}
c = \sum^{\infty}_{k=0}p_{\rm deg}(k)\:k
\end{equation} 
and the second moment $\overline{J^2}=\int \mathrm{d}x~p_J(x)x^2$ of the bond disorder. Eq. \eqref{finalformula} is one of the main results of this paper.

From Eq. \eqref{finalformula}, we observe that $S(t)$ is universal in the sense that it only depends on $p_J(x)$ and $p_{\rm deg}(k)$  through the single parameter $r$.        On the other hand, the dependence on $p_D$ is explicit.   

The present section is organized as follows.   In the first subsection, we discuss the spectral properties of  adjacency matrices of directed random graphs with a prescribed joint degree distribution.   In the second subsection, we derive the formula \eqref{finalformula}.   In the third and fourth subsections,  we derive  explicit expressions for $S(t)$ by computing  the contour integral in Eq. \eqref{finalformula} for fixed diagonal disorder 
\begin{equation}
p_D(x) = \delta(x+\mu)\ , \label{eq:fixed}
\end{equation}
and for  bimodal diagonal disorder, 
\begin{equation} 
 p_D(x) = (1-q)\delta(x+\mu_1) + q\delta(x+\mu_2)\ ,  \label{eq:bimodal}
  \end{equation} 
respectively.     

\subsection{Spectral properties and asymptotic stability} \label{sec:spec}
 The spectra of  random graphs with a prescribed degree distribution have  been studied in detail in  the limit $N\rightarrow \infty$    in Refs.~\cite{Neri2016, Metz2018, Neri2019}.    If  $c>1$, then the graph contains a giant strongly connected component \cite{Dorogovtsev2001}, which contributes a  deterministic   continuous part to the spectrum and may also contribute deterministic   outliers.      
 
The boundary of the continuous part of the spectrum consists of  all $\lambda_{\rm b}$ solving 
\begin{equation}
r^2\int\mathrm{d}x~p_D(x)\frac{1}{|\lambda_{\rm b}-x|^2}=1\ .\label{boundary}
\end{equation}

The  deterministic outliers 
 $\lambda_{\mathrm{isol}}$  solve the relation
\begin{equation}
c \overline{J}\int\mathrm{d}x~p_D(x)\frac{1}{\lambda_{\mathrm{isol}}-x}=1\ . \label{outlier}
\end{equation} 
 Note that (i) outliers are always real, (ii) if $\overline{J}=0$, there are no outliers, (iii)  the boundary of the continuous part and the location of outlier(s) are universal in the sense that they depend on $p_J(x)$ and $p_{\rm deg}(k)$ only through a single parameter (either $r$ or $c \overline{J}$).

\subsection{General diagonal disorder} 

\begin{figure*}[t!]
\centering
\includegraphics[width=\textwidth]{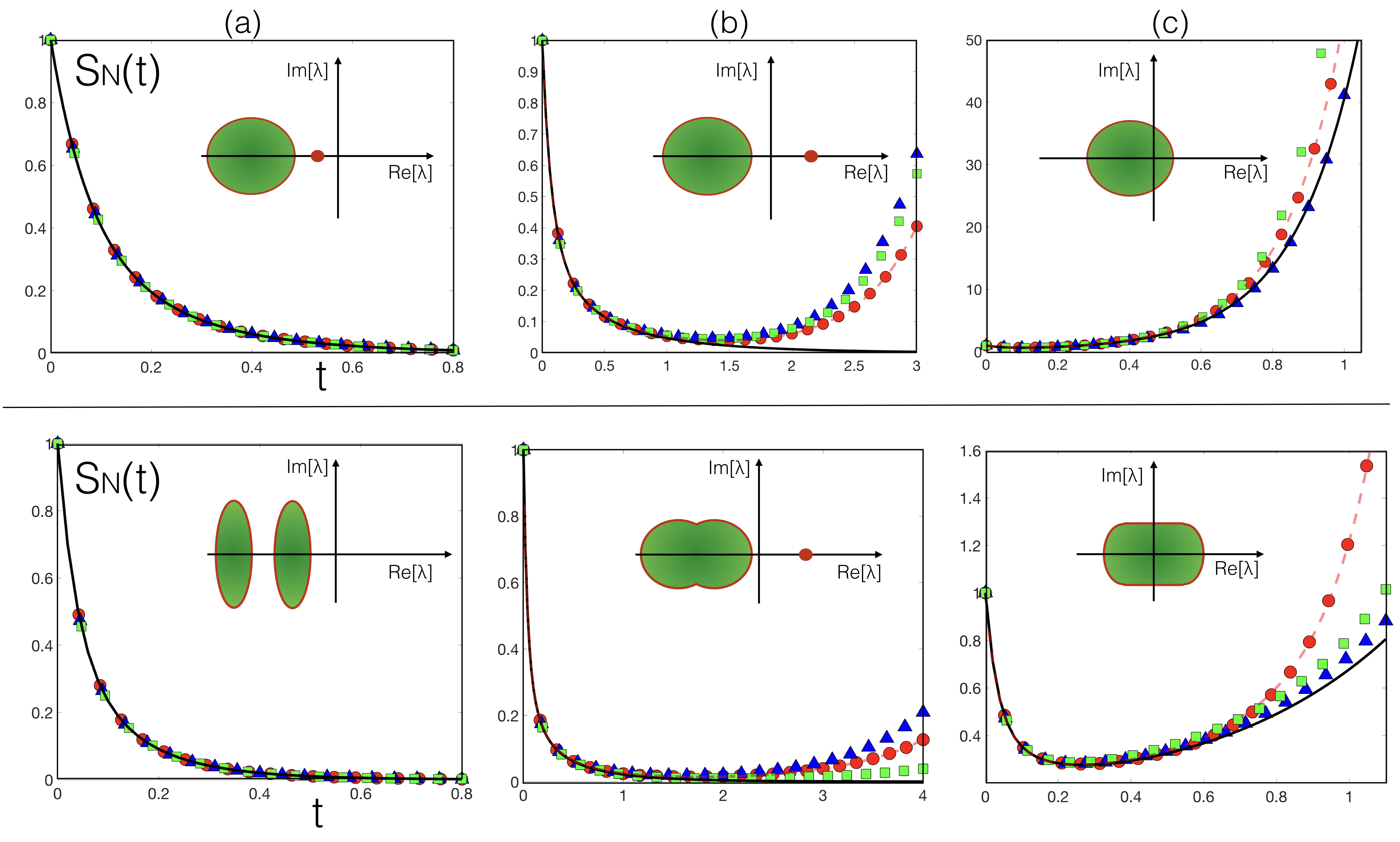}
\caption{
$S_N(t) =\overline{\langle |\bm y(t)|^2\rangle} $ for weighted oriented graphs with Poissonian connectivity with mean degree $c=2$, with fixed diagonal (Top Row) and bimodal diagonal disorder (Bottom Row). The theoretical result for $S(t)$ is provided in black solid line (see Eqs. \eqref{besseldiag} and \eqref{StBimodal}, respectively). Symbols denote numerical solution of the differential equation \eqref{dynamical1} for $N=5000$, averaged over $25$ initial conditions and $5$ realizations of the underlying graph. Red circles stand for Gaussian bond disorder, blue triangles for uniform bond disorder, and green squares for Laplace-distributed disorder. The parameters for different panels are described below. We show schematically in the insets the location of the continuous part of the spectrum and the outlier (if present), according to Eqs.  \eqref{boundary} and \eqref{outlier}.
{\bf Top row:}  Fixed diagonal at $-\mu=-5$. (a) $\overline{J}=2$ and $\overline{J^2}=5$, (b) $\overline{J}=3$ and $\overline{J^2}=10$, and (c) $\overline{J}=4$ and $\overline{J^2}=32$. In panels (b) and (c), the red dashed curves represent $\tilde{S}(t)$ with fitted values of parameters $a=7.7\cdot 10^{-4}$, $b=1.04$  and $a=4.2\cdot 10^{-3}$, $b=4.03$, respectively. Eq. \eqref{boundary} simplifies in this case as $r^2=c\overline{J^2}=|\lambda_{\rm b}-\mu|^2$ and $\lambda_{\mathrm{isol}}=c\overline{J}-\mu$. {\bf Bottom row:} Diagonal entries taken at random between $-\mu_1=-5$ and $-\mu_2=-14$ with equal probability ($q=1/2$).  (a) $\overline{J}=2$ and $\overline{J^2}=5$, (b) $\overline{J}=4$ and $\overline{J^2}=17$, and (c) $\overline{J}=4$ and $\overline{J^2}=32$. 
In panels (b) and (c), the red dashed curves represent $\tilde{S}(t)$ with fitted values of parameters $a=1.1\cdot 10^{-3}$, $b=0.6$  and $a=1.8\cdot 10^{-4}$, $b=4.0$, respectively.
 }\label{Fig1}
\end{figure*}

Since directed random graphs with a prescribed degree distribution are locally tree-like and locally oriented, the Eqs. \eqref{alphajk}-\eqref{alphak} apply.    
We  obtain an analytical expression for the average 
\begin{equation}
\overline{\alpha_k}= \overline{[\mathsf{G}_k]_{1,1}} = \overline{\alpha}
\end{equation}
by taking the ensemble average of Eq. \eqref{alphajk}.  

Using that  (i) the right-hand-side of Eq. \eqref{alphajk} is identical to the right-hand-side of Eq. \eqref{alphak} for any $j,k$ with $C_{jk} = 1$; (ii) all variables on the   r.h.s. of Eq. \eqref{alphajk} are  statistically independent, as degree-degree correlations are absent;  and (iii) all nodes in the ensemble are statistical equivalent, we obtain   
the equation 
\begin{equation}
\overline{\alpha}=(r^2\overline{\alpha}-1)\int \mathrm{d}x~p_D(x)\frac{1}{(z-x)(w-x)}\ .\label{mainalpha}
\end{equation}

Solving \eqref{mainalpha} for $\overline{\alpha}$ and using 
\begin{equation}
\overline{\mathcal{W}_A(z,w)}=-\overline{\sum_{j=1}^N [\mathsf{G}_j]_{1,1}}= -N\overline\alpha\ ,
\end{equation}
we obtain the general formula Eq. \eqref{finalformula} for $S(t)$.     

It remains to perform the contour integral in \eqref{finalformula} for some specific choice of the diagonal disorder, which is the subject of the next subsections.

\subsection{Fixed diagonal disorder}
We compute the contour integral in Eq. \eqref{finalformula} in the case of fixed disorder given by Eq. \eqref{eq:fixed}. 
The integral \eqref{finalformula} can be performed using residues. Changing variables $z'=z+\mu$ and $w'=w+\mu$, we obtain
\begin{align}
 S(t) 
\! &=\frac{1}{2\pi\mathrm{i}}e^{-2\mu t}\oint_{\gamma'}\mathrm{d}z' \frac{e^{t(z'+r^2/z')}}{z'}\!  \nonumber\\
&=e^{-2\mu t}I_0(2rt)\ .\label{besseldiag}
\end{align}
Quite remarkably, Eq. \eqref{besseldiag} for sparse oriented graphs and Eq. \eqref{Sfullyconnected} for fully connected structures share the same functional form, and therefore the two models fall into the same universality class.  Indeed, if $\overline{J}=0$, then $r$ is the spectral radius $\rho$ of $A$ (see \eqref{boundary}). 

For  large $t$,
\begin{equation}
S(t)\sim \frac{1}{\sqrt{4\pi r t}}e^{2t(r-\mu)}\ , \label{eq:22}
\end{equation}
where the exponent describes the asymptotic growth or decay at a rate $r-\mu$.   The rate  $r-\mu$ contains a positive contribution $r$, which is the amplification of the initial perturbation when  it spreads throughout the network, and a negative contribution $-\mu$, which is the local decay rate.

For  $t\approx 0$, it holds that  
\begin{equation}
S(t) = [1 - 2\mu t+\mathcal{O}(t^2)]\ , \label{eq:init}
\end{equation}
independent of the network structure.   

While the initial response is independent of the network structure and leads to a decrease in $S(t)$, the response at larger $t$ will depend on how the initial perturbation  spreads through  the network.    In particular, if $r-\mu>0$, then the    response  $S(t)$ is non-monotonic.

It is insightful to interpret the result given by Eq. \eqref{eq:22}  in the light of  the spectral properties of the ensemble, as discussed in Sec.~\ref{sec:spec}.   The boundary of the continuous part of the spectrum is according to Eq. \eqref{boundary} a circle of radius $r$ centered around $-\mu$.   As a consequence, the eigenvalue with the maximum real part, which belongs to the continuous spectrum, is given by 
\begin{equation}
\lambda'_{\rm b} = \max_{\lambda_{\rm b}} \mathrm{Re}[\lambda_{\rm b}] = r-\mu\ .
\end{equation}
This implies that the asymptotic dynamics of $S(t)$, given by  Eq. \eqref{eq:22} for large $t$,  reads 
\begin{equation}
S(t)\sim \frac{1}{\sqrt{4\pi r t}}e^{2t \lambda'_{\rm b}}\ ,\label{eq:22xx}
\end{equation} 
and the asymptotic rate is  governed by the boundary of the continuous part of the spectrum.    The outlier 
\begin{equation}
\lambda_{\rm isol} =  c\overline{J}
\end{equation}
plays no role in the behavior of $S(t)$, even when $\lambda_{\rm isol}>\lambda'_{\rm b}$.    This is because the initial perturbation is random and  therefore -- in the limit $N\rightarrow \infty$ -- it stands orthogonal to the one-dimensional eigenspace spanned by the outlier.

\subsection{Bimodal diagonal disorder}
In the case of bimodal diagonal disorder with probability density $p_D$ given in Eq. \eqref{eq:bimodal}, we evaluate the integral in Eq. \eqref{finalformula} using residues.  After some calculations, presented in  Appendix \ref{appB}, we obtain
\begin{align}
\nonumber S(t) &=(1-q)e^{-2\mu_1 t}I_0(2 r t \sqrt{1-q})\\
&+qe^{-2\mu_2 t}I_0(2 r t \sqrt{q}) +e^{-2\mu_1 t}\Psi(t)\ ,\label{StBimodal}
\end{align}
where  
\begin{align}
\nonumber \Psi(t) &=\sum_{m\geq 1}\frac{(rt)^{2m}}{(m!)^2}\sum_{n=1}^m  \binom{m+1}{n} q^n(1-q)^{m-n+1}\\
&\times~ [_1F_1(n;m+1;-(\mu_2-\mu_1) t)]^2\ , \label{StBimodalT}
\end{align}
is a series involving confluent hypergeometric functions $_1F_1$  \cite{dlmf}.      The confluent hypergeometric function is defined by the series 
   \begin{equation}
  _1F_1(a;b;z) = \sum^{\infty}_{n=0}\frac{(a+n-1)^{\underline{n}}}{(b+n-1)^{\underline{n}}}\frac{z^n}{n!}, \label{eq:Kummer}
\end{equation}
where 
\begin{equation}
(a)^{\underline{s}}=\prod_{j=0}^{s-1}(a-j)\ , \quad  s\in \mathbb{N}\ ,  \label{eq:falling}
\end{equation}
 is the falling factorial and $(a)^{\underline{0}}= 1$. Although not immediately evident, the formula \eqref{StBimodal} is symmetric in $\mu_1$ and $\mu_2$ upon the exchange $q\to 1-q$ as it should, due to a transformation formula of the confluent hypergeometric function upon change of sign of the main argument.

The formula \eqref{StBimodal} is  rather   complicated, but it has an appealing dynamical interpretation.  In this example, the dynamical system consists of two sub-populations with decay rates $\mu_1$ and $\mu_2$, respectively. Neglecting interactions between the subpopulations, each of these would evolve in isolation according to Eq.~\eqref{besseldiag}, albeit with reduced connectivities $qc$ and $(1-q)c$, respectively. The first two terms in \eqref{StBimodal} describe precisely the dynamics of the populations in isolation. The third term instead describes the dynamical interference  between the two sub-populations.   We expect that the interference will be important in the limit of large $t$.

For  $t\approx 0$, it holds that  
\begin{equation}
S(t) = [1 - 2(1-q)\mu_1 t - 2q \mu_2 t +\mathcal{O}(t^2)]\ , \label{eq:init2}
\end{equation}
which is again independent of the network structure.   

The analysis of the $t\rightarrow \infty$ limit of Eq. \eqref{StBimodal} is  more complicated because of the nontrivial form of the interference term.   Nevertheless, based on the analysis in the previous subsection for  fixed diagonal disorder, we expect that for large $t$
\begin{equation}
S(t)  \sim  e^{\lambda'_{\rm b}t}
\end{equation}
where 
\begin{equation}
\lambda'_{\rm b} =  \max_{\lambda_{\rm b}} \mathrm{Re}[\lambda_{\rm b}]
\end{equation}
is the value of   $\lambda_b$, located at the boundary of the continuous part of the spectrum, with the largest real part.     In the present example with $p_D$ the sum of two Dirac distributions, we obtain (see Appendix~\ref{bimodalSpec})
\begin{equation}
\lambda'_{\rm b} =    \frac{-\mu_1-\mu_2 + \sqrt{(\mu_1-\mu_2)^2+4r^2 q +4\sqrt{\mathcal{D}}}}{2}\ ,
\end{equation}
where 
  \begin{equation}
    \mathcal{D} = r^2   \mu_1^2 q - 2 r^2  \mu_1 \mu_2 q +r^2 \mu_2^2 q + r^4  q^2\ .
  \end{equation}  
Hence,  for bimodal diagonal disorder even  the asymptotic rate $\lambda'_{\rm b}$ does not admit a simple expression, which clarifies why it is not a simple task to analyze the Eq. \eqref{StBimodal} in the limit of large $t$.  

\section{Numerical experiments on random matrices and  on empirical networks}\label{sec:Empirical}  
We compare the obtained analytical expression for $S(t) = \lim_{N\rightarrow\infty} S_N(t)$, given by Eq. \eqref{besseldiag} and  Eqs. \eqref{StBimodal}-\eqref{StBimodalT}, with numerical results for $S_N(t)$  on random matrices and on empirical networks.

\subsection{Random graphs}
We analyze  the dynamics described by Eq. \eqref{dynamical1} with $A$ being the adjacency matrix of a  weighted random graph with a prescribed degree distribution (see the Introduction for the definition of this ensemble).      Here, we use a Poissonian
distribution for the  in-degrees (and out-degrees), namely, 
\begin{equation}
p_{\mathrm{deg}}(k) =  \frac{e^{-c}c^{k}}{k!}
\end{equation}
 in Eq. \eqref{eq:prescr}.   In addition, in Appendix~\ref{powerlaw}, we show results for power-law degree distributions, which are relevant to describe real-world networks, see Refs.~\cite{Newman2010, Clauset2009, Dorogovtsev2008, Broido, Voitalov}.

 In Fig.~\ref{Fig1}, we compare the obtained  expressions for $S(t)$ with estimates of $S_N(t)$ obtained from numerical experiments.    We consider three qualitatively different scenarios: (a) the system is both transiently and asymptotically stable ($\lambda'_{{\rm b}}<0$ and $\lambda_{\mathrm{isol}}<0$), (b) the system is transiently stable but asymptotically unstable ($\lambda'_{{\rm b}}<0$ and $\lambda_{\mathrm{isol}}>0$), and (c) the system is unstable ($\lambda'_{{\rm b}}>0$).  
Since the theory in Eq. \eqref{finalformula} is obtained by taking the limit $N\to\infty$ at \emph{fixed} time, while in the simulations we work with a fixed system size $N$ and look at its time evolution, there will be a crossover time $t^\star(N)$ after which the theory and simulations are expected to diverge (see  Appendix~\ref{App:NdependN} for a detailed discussion).  
However, for $t<t^\star(N)$, we observe perfect agreement between theory and simulations, proving that the precise connectivity structure and the type of bond disorder do not matter in the transient dynamics of large dynamical systems.  Only the combination $r^2=c\overline{J^2}$ plays a role, as predicted by the theory. In case (a), theory and simulations are in very good agreement, and $1/(2(r-\mu))$ provides the typical relaxation time to stationarity.

  We clearly see the effect of the crossover time $t^\star(N)$ in cases (b) and (c) of Fig. \ref{Fig1} [but this would also become apparent in case (a) when $t$ is large enough and if a different scale were used].      For $t\gg t^\star(N)$, the contribution from the eigenmode with largest -- and positive -- real part $S_N(t;A)\sim a_N e^{2t\mathrm{Re}[\lambda_1]}$ (where $a_N$ is independent of $t$ and vanishes for large $N$) will govern the large-$t$ behavior of every single realization of the $N\times N$ numerical experiment, and as a consequence also of the average  $S_N(t) = \overline{S_N(t;A)}$. 
  
 In order to capture both the transient and asymptotic behaviors, we can modify $S(t)$ by including the effect of the largest eigenmode as
\begin{equation}
\tilde{S}(t)=S(t)+a e^{2bt}\ ,\label{fitS}
\end{equation}
where $a$ and $b$ are two (albeit non-universal) fitting parameters. This is illustrated in Fig. \ref{Fig1} (b) and (c). We observe empirically that universality is broken after $t^\star(N)$: this is because the exponential growth of $S_N(t;A)$ at large times -- for every individual realization of the matrix $A$ -- is very sensitive to the \emph{precise} value of $\lambda_1(A)$, which of course fluctuates from one realization to another. The average $S_N(t) = \overline{S_N(t;A)}$ is therefore disproportionally affected by the \emph{largest} $\lambda_1(A)$ across the sample used for the simulations.

Note also that in the panel (b) the fitted exponent $b\approx\lambda_{\mathrm{isol}}$, whereas in  panel (c) the exponent $b$ is different from $\max\mathrm{Re} [\lambda_{\rm b}]$.

\subsection{Empirical networks} 
We test  how well Eq. \eqref{besseldiag}  describes the transient dynamics of a dynamical system defined on a  real-world graph.   To this aim, we numerically solve the dynamics in Eq. \eqref{dynamical1} for a matrix of the form Eq. \eqref{eq:modelDef},  with  $C_{ij}$  the entries of the adjacency matrix of a real world network.   We consider two examples of real-world networks, namely, a food web and a signaling network.  

For the food web example, we have chosen the food web of Otago Harbour,  an intertidal mudflat ecosystem in New Zealand, whose adjacency matrix is determined in Ref.~\cite{Mouritsen}.  This adjacency matrix contains  180 nodes and 1924 edges.   Nodes represent species, which include among others, plants, annelids, birds, and fish.     Links represent trophic interactions between species, including among others,  predation, parasitic castration, and macroparasitism interactions. The food web of Otago Harbour is an (almost) oriented network: $97\%$ of the links are oriented.    The mean out-degree is $c = 1924/180 \approx 10.7$.

For the example of a cellular signaling network, we have considered  a network of molecular interactions between signaling proteins in human cells.    We have collected  data from the NCI-Nature Pathway Interaction database
 \cite{Schaefer}, which is now available in the Network Data Exchange, NDEx \cite{ Pratt1, Pillich, Pratt2}.   We have extracted the adjacency matrix formed from the nodes in the  2-step neighborhood  of the protein kinases MAPK1, AKT1, JAK1 and the protein APC.    The resultant adjacency matrix contains 2288 nodes and 23207 edges.   The mean out-degree is  $c \approx  13.3$ and  $70\%$ of the edges are oriented.

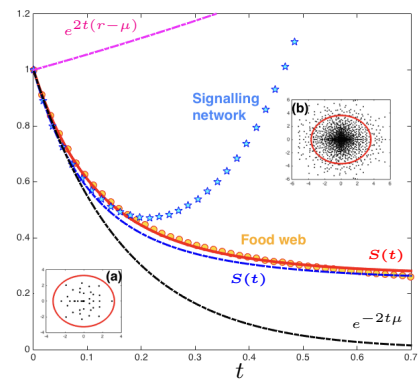
\begin{figure}[t!]
\centering
\includegraphics[width=0.3\textwidth]{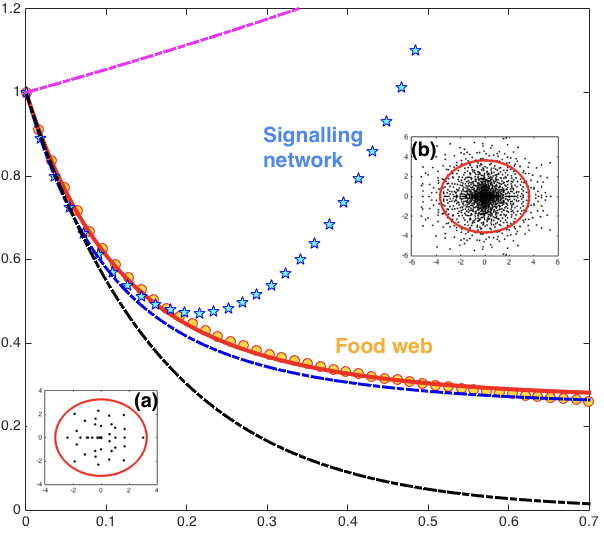}
 \put(-135,125){\color{magenta}\tiny$e^{ 2t(r-\mu)}$}
  \put(-26,14){\tiny$e^{-2 t\mu}$}
  \put(-20,40){\color{red}\tiny \mbox{\boldmath$S(t)$}}
    \put(-72,31){\color{blue}\tiny \mbox{\boldmath$S(t)$}}
        \put(-70,-5){$t$}
\caption{$S_N(t)=\overline{\langle |\bm y(t)|^2\rangle}$ for the simulated dynamics  of Eq.~\eqref{dynamical1} on two examples of real-world graphs (circles  for a food web with $N=180$ and $c=10.7$, and stars  for a signaling network with $N=2288$ and $c=13.3$) are compared with the theoretical prediction $S(t)$ (blue and red lines), given by Eq. \eqref{besseldiag}, as well as with, $e^{2t(r-\mu)}$ and $e^{-2\mu t}$ (black and magenta lines).    We have weighted the networks with couplings $J_{ij}$  that are   i.i.d.~random variables drawn from the distribution $p_J(x) = (1/2) \delta(x-1) +(1/2) \delta(x+1)$ and we have used a single  realization of the matrix $J$ for both the food web and the signaling network.     The diagonal elements are fixed to a constant   $-\mu = -r+0.27$,  such that the system is asymptotically unstable.     Insets (a)  and (b) show the spectra of the  adjacency matrices   $A$ for the  food web and the signaling network considered with random couplings.   The red circle has radius  $r = \sqrt{c}$  and is the predicted  boundary of the spectrum according to \eqref{boundary}.     We have estimated $S_N(t)=\overline{\langle |\bm y(t)|^2\rangle}$ by simulating the dynamics on the generated networks for $25$ realizations of the initial condition $\bm y(0)$.
 }\label{Fig2}
\end{figure}

In Fig.~\ref{Fig2}  we present simulated dynamics on these networks.    Numerical results show that the  theory  describes very well  the transient dynamics on the food web (for all $t<t^\star(N)$), in spite of its small size $N=180$, while the dynamics of the signaling network is less well captured.  Plotting the full spectra (see insets of Fig.~\ref{Fig2}), we see that the largest eigenvalue of the food web is almost equal to $\sqrt{c}$ (predicted by Eq. \eqref{boundary}), whereas for the signaling network it is much larger than $\sqrt{c}$, which clarifies why the theory works better for the food web.

In order to further demonstrate the relevance of Eq. \eqref{besseldiag} in modeling dynamics on real networks, we compare $S_N(t)$ with a naive theory based on either an  exponential decay $e^{-2 t\mu}$ or on exponential growth  $e^{2t(r - \mu)}$.   We observe that the naive theories do not capture well the  transient response of the dynamical system.     This is because the response of the real system consists of an initial decay at rate $\mu$ and asymptotic growth at rate $r - \mu$, which govern how the initial random perturbation propagates through the networks.  

\section{Conclusions and Outlook}\label{Disc} 
We have developed a mathematical formalism that allows one to study    how  the transient response to an initial perturbation of a large dynamical system, captured by the observable $S_N(t)$, depends  on the topology of the underlying network of interactions.   The developed method allows one to compute the limiting value $S(t) = \lim_{N\rightarrow \infty}S_N(t)$ for graphs that have a     locally tree-like structure.    As an example, we have studied the transient dynamics on directed random graphs with prescribed degree distribution, which are canonical models for real-worlds systems, such as, the Internet \cite{Broder, Dorogovtsev2001, Pastor},  neural networks \cite{Brunel} and other high-dimensional systems \cite{Newman2001, Dorogov2013, Newman2010}.  We have found that the transient response of large systems is universal, in the sense that $S(t)$ only  depends on the network topology through the   single parameter $r^2 = c\overline{J^2}$, which is the product of the mean degree and the second moment of the distribution of interactions strengths. 
   On the other hand, the dependence on the statistics of diagonal elements, given by the distribution $p_D$, is non-universal.  

The developed method is fairly general, and therefore various other examples can be considered.   For example, it should be straightforward to extend the presented results to random graphs with correlations between in-degrees and out-degrees \cite{Neri2019} or to non-oriented systems defined on regular graphs \cite{Neri2012}.   More challenging, but still feasible,  is to extend the theory to graphs that contain many small cycles  \cite{Metz2011,Bolle2013, Aceituno}.  

From a random matrix theory point of view, in the present paper we have developed the mathematical theory for the 2-point correlator  $\mathcal{W}_A(w,z)$ of sparse random graphs.   We can extend the present formalism to account for  higher order $2n$-correlators, which provide information on higher-order moments of $S_N$ \cite{TarnowskiPrep}.

Since the theoretical response functions,  given by Eqs. \eqref{besseldiag} and  \eqref{StBimodal},  only depend on a few parameters and describe well the transient response of dynamical systems defined on real-world networks, we believe that  Eqs. \eqref{besseldiag} and  \eqref{StBimodal} can be used to infer properties of networks from time-series data. 

\begin{acknowledgments}
WT is supported by ETIUDA scholarship 2018/28/T/ST1/00470 from National Science Center and the Diamond Grant 0225/DIA/2015/44 from the Polish Ministry of Science and Higher Education. WT is grateful to King's College London for hospitality, where this work was done. PV and IN acknowledge funding by the Engineering and Physical Sciences Research Council (EPSRC) through the Centre for Doctoral Training in Cross Disciplinary Approaches to Non-Equilibrium Systems (CANES, Grant Nr. EP/L015854/1).   We thank Ahmed Zaid for making us aware of a  sign typo in some equations of the paper. 
\end{acknowledgments}

\appendix

\section{$N$-dependence of the transition time $t^\star(N)$ }\label{App:NdependN}
We study here how the  crossover  time  $t^\star(N)$ depends on $N$.   The crossover time determines when  the system changes from its   transient regime, where $S_N(t) = \overline{\langle |\mathbf{y}(t)|^2\rangle} \approx S(t) = \lim_{N\rightarrow\infty}S_N(t)$, to the asymptotic regime, where $S(t) \ll  S_N(t)$.    
For transiently stable systems,   we define the crossover time $t^\star(N)$ by Eq. \eqref{tstardef}.

In the first subsection, we determine the $N$-dependence of  $t^\star(N)$ with a brute-force numerical analysis on  four random matrix ensembles.  In the second subsection,  we analyze 
$t^\star(N)$ in the case of the normal Ginibre ensemble, a case for which we have the rare luxury of being able to carry out the  analytical treatment in full.  We obtain an explicit analytical expression for $S_N(t)$, which can be used to evaluate numerically its minimum value.  We find that  $t^\star(N) \approx  1.15 \sqrt{N}$.

\subsection{Numerical results for four ensembles}\label{tstar}
We study the first-order dynamics in Eq.~\eqref{dynamical1} for 
\begin{equation}
A_{ij}=-\mu\delta_{ij}+X_{ij}\ . \label{eq:x}
\end{equation}

\begin{figure}
\includegraphics[width=0.45\textwidth]{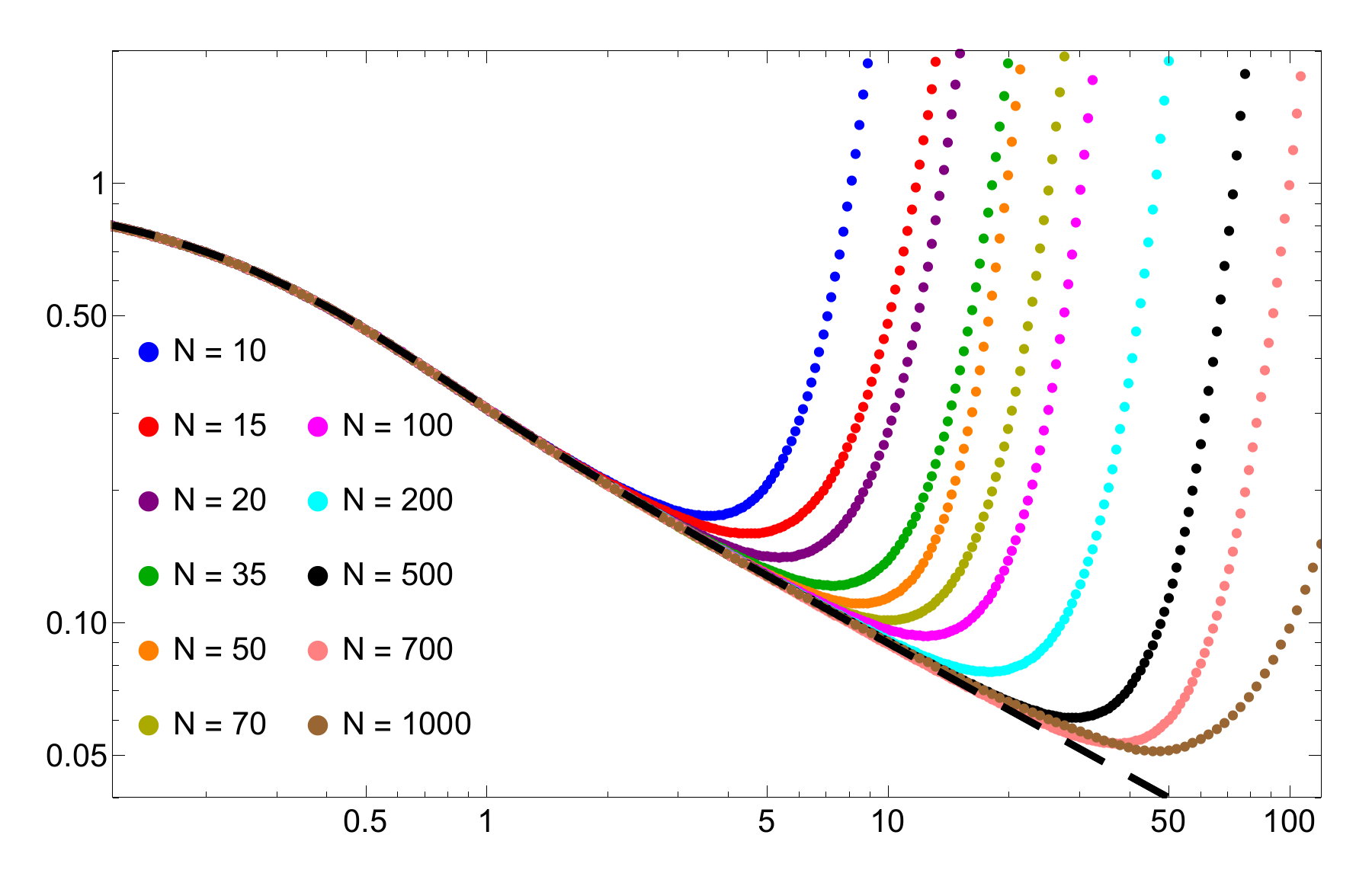}
  \put(-150,0){$t$}
\caption{Markers denote numerical results  for $S_N(t) =\overline{\langle |\bm y(t)|^2\rangle}$  for model \eqref{dynamical1} with matrix \eqref{eq:x} as a function of $t$  and for different values of $N$.      The $X_{ij}$ are i.i.d.~random variables taken from a   Gaussian distribution with  zero mean and variance $1/N$  and we have set $\mu=-1$.    Dashed black line is the curve $S(t)=e^{-2\mu t}I_0(2\rho t)$ (see Eq. \eqref{Sfullyconnected}).   Markers are averages over   $2000$ matrix samples  with $25$  realizations of the initial conditions for each sample.   
\label{Fig3}
}
\end{figure}

In Fig.~\ref{Fig3} we present numerical results for    $S_N(t)$ as a function of  $t$ for different values of $N$ in the case of  $X_{ij}$ being i.i.d.~ random variables taken from a Gaussian distribution.   
Fig.~\ref{Fig3} shows that $S(t)\approx S_N(t)$ for small enough  $t$  (transient regime) while $S(t)\ll S_N(t)$ when $t$ is large enough (asymptotic regime).   Moreover, we observe that the crossover time $t^\star(N)$ from the transient regime  to the asymptotic regime grows as a function of $N$.  

\begin{figure*}[t!]
\centering
\includegraphics[width=0.99\textwidth]{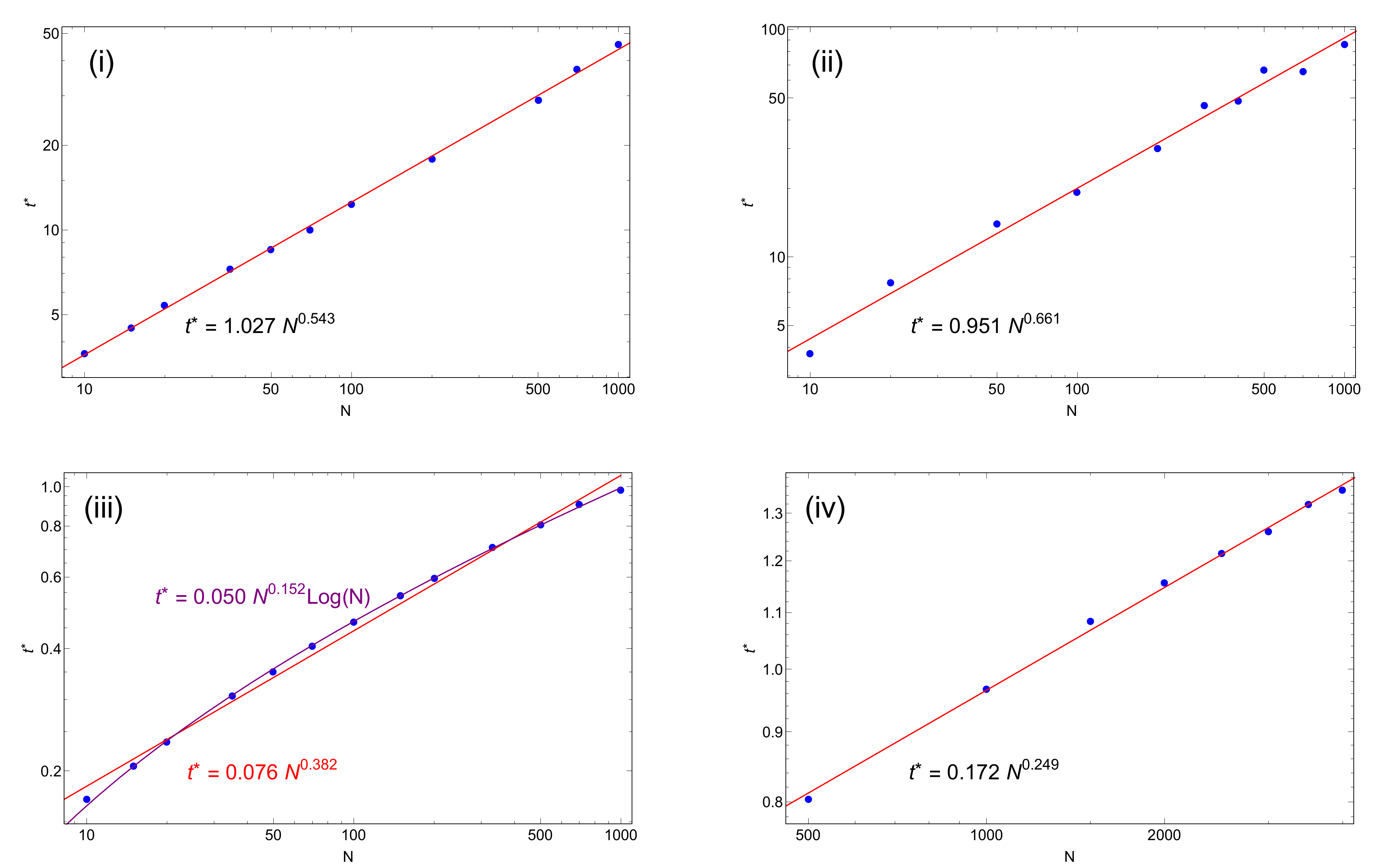}
\caption{ Numerical results for the crossover time $t^\star$ as a function of $N$ for the four considered models: (i) Ginibre ensemble, (ii) Gaussian Orthogonal Ensemble, (iii) Ginibre ensemble with an outlier, (iv) sparse random graph with Poissonian connectivity, mean degree $c=2$ and Gaussian bond disorder. Solid lines denote fitted functions to the empirical data.  
In cases (i) and (ii) we have set $\mu$ such that the system is at the edge of stability (in the $N\to\infty$ limit, the average of the leading eigenvalue $\overline{\lambda_1}=0$), and therefore  $S(t)$ decays asymptotically as $t^{-1/2}$. 
The parameters in  (iii) and (iv) are  set such that the system is transiently stable but asymptotically unstable, as in the  case (b) of the top panel of Fig.~\ref{Fig1}.
In cases (i)-(iii), markers are  averages over   $2000$ matrix samples  with $25$  realizations of the initial conditions each.     In  case (iv), we have used $20$ matrix samples with   $25$ realizations each.    
\label{Fig2Supp}
}
\end{figure*}

In Fig.~\ref{Fig2Supp} we analyse how the crossover time $t^\star(N)$ depends on $N$.  For this we use its definition given by Eq. \eqref{tstardef}.   We consider four random matrix ensembles: 
 \begin{enumerate}[label=(\roman*)]
 \item Ginibre matrices with zero mean \cite{Ginibre1965}: the $X_{ij}$ are i.i.d.~random variables taken from a   Gaussian distribution with  zero mean and variance $1/N$,  and we set $\mu=-1$; 
 \item Gaussian Orthogonal Ensemble:   the $X_{ij}=X_{ji}$ are i.i.d.~real-valued random variables taken from a   Gaussian distribution with  zero mean and variance $1/N$, and we set $\mu=2$; 
 \item Ginibre matrices with nonzero mean: the $X_{ij}$ are the  i.i.d.random variables   taken from a   Gaussian distribution with   mean $6/N$ and variance  $20/N$, and we set $\mu=5$; 
 \item Adjacency matrices of  sparse random  graphs:    we set $X_{ij} = J_{ij}C_{ij}$ with $C_{ij}$ the adjacency matrix  of a random graph  with a  prescribed  joint  distribution of in-degrees and out-degrees given by 
 \begin{equation}
 p_{\mathrm{deg}}(k_{\mathrm{in}},k_\mathrm{out}) = \frac{e^{-c}c^{k_{\rm in}}}{k_{\rm in}!}\frac{e^{-c}c^{k_{\rm out}}}{k_{\rm out}!}, 
 \end{equation}
  with a mean in-degree (out-degree)  $c=2$.  In addition, we  weigh the edges with  couplings $J_{ij}$ that are i.i.d.~random variables taken from a Gaussian distribution with mean $3$ and variance~$1$.  We set $\mu=5$.   
 \end{enumerate}

   In cases (i), (ii) and (iv) the numerical results for $t^\star$  are well fitted by the power law $t^\star=\alpha N^{\beta}$.  In the case (iii)  a pure power-law curve does not describe the data as well as the function  $t^\star=\alpha N^{\beta}\log N$, i.e., a power-law with a logarithmic correction.  

Interestingly, for the Ginibre ensemble and GOE (cases (i) and (ii)) the fitted exponents are approximately $1/2$ and $2/3$, respectively, which is the exponent governing the scale of typical fluctuations at the edge of the spectrum \cite{Rider, Bowick}. This pattern, however, does not seem to carry over to cases (iii) and (iv).    We are not aware of a theory that can shed more light on  the scaling of $t^\star$ with $N$ in these cases.

In conclusion, numerical results  indicate that $t^\star$ diverges with $N$ with a law that might be related to the scale of typical fluctuations at the edge of the spectrum.     In the next section, we analyze in more detail the normal Ginibre ensemble, which allows for a more complete analytical treatment.

\subsection{Analytical treatment  of $t^\star(N)$ for the normal Ginibre ensemble}\label{normalgin}
We analyze the $N$-dependence of  $t^\star(N)$ for a rare example where a full analytical treatment is possible, namely the normal Ginibre ensemble.  For this ensemble, we  derive an explicit analytical expression for $S_N(t)$.  Evaluating numerically its minimum, we obtain that  
\begin{equation}
t^\star(N) \approx 1.15 \sqrt{N}\ .
\end{equation}

\subsubsection{Definition of the normal Ginibre ensemble} 
We consider dynamical systems of the type given by Eq. \eqref{dynamical1} with 
\begin{equation}
A=-\mu\mathds{1}+X'/\sqrt{N}\ ,
\end{equation}
where $X'$ is a matrix drawn from the normal Ginibre ensemble. 

Normal Ginibre matrices are obtained by retaining the normal term in the Schur decomposition of a complex Ginibre matrix. 
The  entries of the complex Ginibre ensemble \cite{Ginibre1965} are defined as  follows, 
\begin{equation}
X_{jk} =x_{jk}+\mathrm{i}y_{jk}\ ,
\end{equation}
where $x_{jk}$ and $y_{jk}$ are i.i.d. random variables from  a Gaussian distribution  with zero mean and variance $1/2$. The matrices in the ensemble are  non-symmetric. 
The joint pdf of matrix entries reads
\begin{equation}
P_X(X)=C_N \exp\left[-\mathrm{Tr}(XX^{\dagger})\right], \label{eq:PX}
\end{equation}
where $C_N$  is the normalization constant.  

In order to obtain a normal Ginibre ensemble, we implement 
the
Schur decomposition 
\begin{equation}
X=U(\Lambda+T)U^{\dagger} \label{eq:normGin}
\end{equation}
with  $U$ a unitary matrix, $\Lambda$  the diagonal matrix of eigenvalues, and $T$ a strictly upper triangular matrix.    From Eq. \eqref{eq:normGin} it is apparent that the upper-triangular matrix  $T$  is the source of non-normality.   Therefore, we drop the upper-triangular matrix to obtain 
\begin{align}
X' = U\Lambda U^{\dagger}\ ,
\end{align}
which defines a matrix sample drawn from the normal Ginibre ensemble.  

Implementing the Schur decomposition in Eq. \eqref{eq:PX}, we find that $P_X$ factorizes into 
\begin{equation}
P_X(X)=P_T(T)P_{\Lambda}(\Lambda)P_{U}(U)\ , \label{eq:PXx}
\end{equation}
where
\begin{equation}
P_{\Lambda}(\Lambda)\sim \exp\left(-\sum_{i=1}^{N}|\lambda_i|^2\right)\prod_{i\neq j}|\lambda_i-\lambda_j|^2
\end{equation}
is the distribution of complex eigenvalues in the \emph{normal} Ginibre ensemble.

\subsubsection{Calculation of $S(t)$}
Because of Eqs. \eqref{eq:SNta}, \eqref{eq:SNTt}, and \eqref{eq:yT}, we can write
\begin{equation}
S_N(t)=\overline{\frac{1}{N}\mathrm{Tr}~ e^{A^{\dagger}t}e^{At}}\ .
\end{equation}
Taking the limit $N\rightarrow \infty$  and using  that the 
eigenvectors of normal (Ginibre) matrices are orthonormal,  we obtain that 
\begin{equation}
S(t)=e^{-2\mu t}\int_{\mathbb{C}^2}\mathrm{d}^2z~ e^{t(z+\zb)}\rho(z,\zb)\ ,\label{eq:Sn}
\end{equation}
where the integration is over the entire complex plane and $\rho(z,\zb)$ is the spectral density of the normal Ginibre ensemble in the limit $N\rightarrow \infty$.    Since in this limit, the spectral density is uniform in the disk of unit radius \cite{Mehta},
\begin{equation}
\rho_{\infty}=\frac{1}{\pi}\idm_{|z|<1}\ ,
\end{equation}
with $\idm_T$  the indicator function (equal to $1$ when $T$ is true and $0$ when $T$ is false), we obtain the expression
\begin{align}
\nonumber S(t) &=\frac{e^{-2\mu t}}{\pi}\int_0^1 r\mathrm{d}r\int_0^{2\pi}\mathrm{d}\varphi~ e^{2rt \cos \varphi}=\\
&=\int_0^{1}2r I_0(2rt) \mathrm{d}r = \frac{e^{-2\mu t}}{t}I_1(2rt)\ ,
\end{align}
where  $I_0$ and $I_1$ are modified Bessel functions. This result  is the analog of Eqs. \eqref{Sfullyconnected} and  Eqs. \eqref{finalformula} for dense and  normal matrices. 

To compute $t^\star$ as a function of $N$, though, the analysis above is not sufficient, and we need to resort to a finite-$N$ calculation.

\subsubsection{Calculation of $S_N(t)$}
For finite $N$, Eq. \eqref{eq:Sn} is modified as
\begin{equation}
S_N(t)=e^{-2\mu t}\int_{\mathbb{C}^2}\mathrm{d}^2z~ e^{t(z+\zb)}\rho_N(z,\zb)\ ,\label{eq:Sn2}
\end{equation}
where $\rho_N(z,\zb)$ is now the spectral density  of the normal Ginibre ensemble at finite  matrix sizes $N$.  It holds that \cite{Chau}
\begin{equation}
\rho_N(z,\zb)=\frac{1}{N\pi} e^{-N|z|^2}\sum_{k=0}^{N-1}\frac{(N|z|^2)^k}{k!}\ .
\end{equation}
Therefore,  we  obtain
\begin{align}
\nonumber S_N(t)&=\frac{e^{-2\mu t}}{\pi}\int_{0}^{\infty}r\mathrm{d}r \int_0^{2\pi}\mathrm{d}\varphi~ e^{2tr\cos\varphi} e^{-Nr^2}\sum_{k=0}^{N-1}\frac{N^k r^{2k}}{k!}  \\
& = 2e^{-2\mu t}\int_0^{\infty}e^{-Nr^2}I_0(2rt)\sum_{k=0}^{N-1}\frac{N^k r^{2k+1}}{k!} \mathrm{d}r\ .
\end{align}
It is convenient to rescale $r\to\frac{s}{\sqrt{N}}$ to get
\begin{equation}
S_N(t)=\frac{2e^{-2\mu t}}{N}\int_0^{\infty}e^{-s^2}I_0\left(\frac{2st}{\sqrt{N}}\right) \sum_{k=0}^{N-1} \frac{s^{2k+1}}{k!}\mathrm{d}s\ .
\end{equation}
The integral can be performed with some work to yield eventually the exact  result
\begin{equation}
S_N(t) = \frac{e^{-2\mu t}}{N} \sum_{k=0}^{N-1} {}_1 F_1\left(k+1;1;\frac{t^2}{N}\right)\ , \label{eq:SNTexact}
\end{equation}
where ${}_1F_1(a;b;z)$ is the  Kummer confluent hypergeometric function, defined in Eq. \eqref{eq:Kummer}.

\subsubsection{Computing the crossover time  $t^\star(N)$}
\begin{figure}[h]
\includegraphics[width=0.45\textwidth]{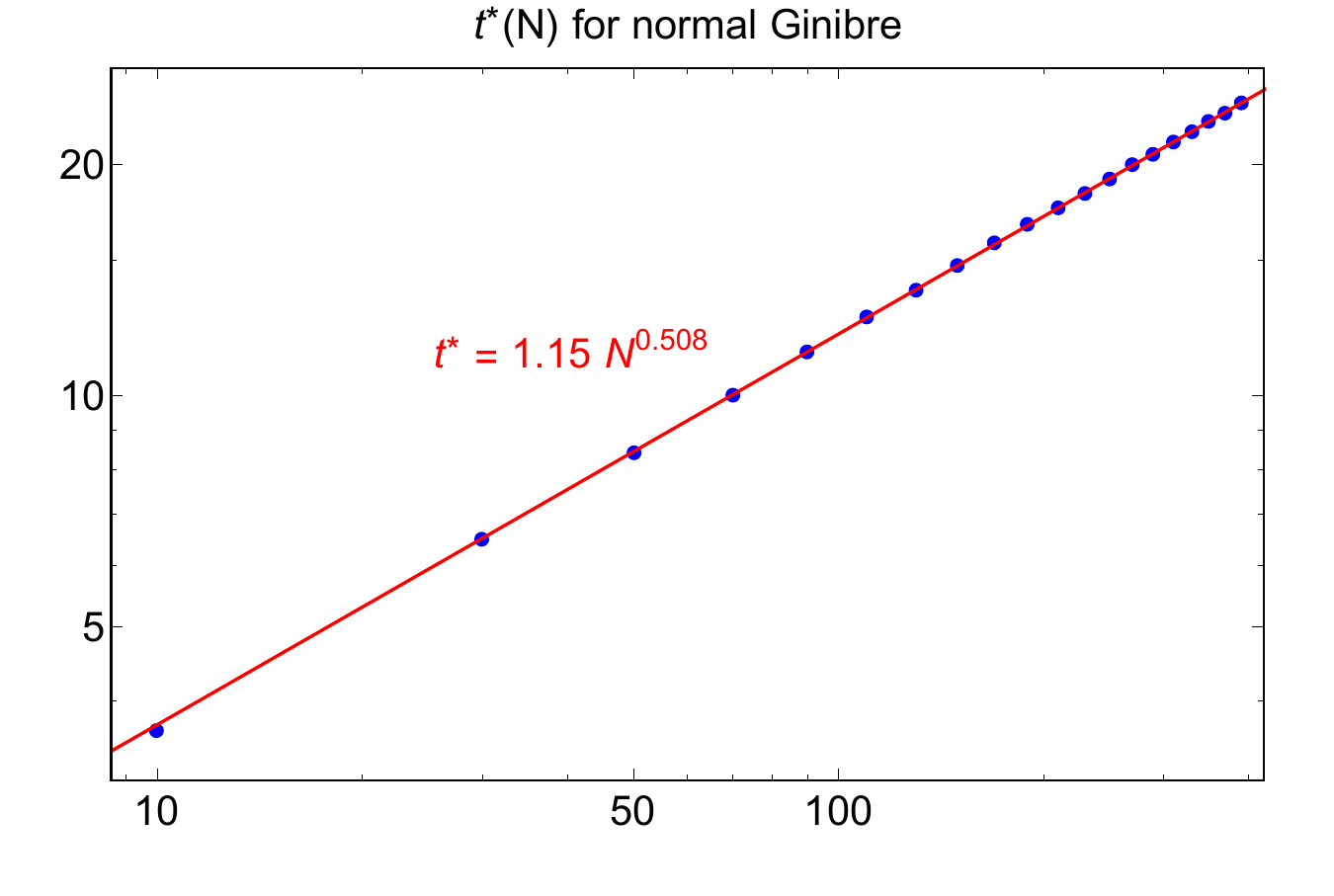}
\caption{Dependence of $t^\star$ on the system size for the  normal Ginibre ensemble, see Appendix~\ref{normalgin}. Points are the numerical solution of  Eq.~\eqref{trans}.  The blue solid line is obtained by fitting the model $t^\star\approx \alpha N^{\beta}$ with 2 free parameters. 
\label{Fig6}
}
\end{figure}

The crossover time $t^\star$ is  defined as the minimum of Eq. \eqref{eq:SNTexact}.   By equating the  derivative of  Eq. \eqref{eq:SNTexact}   to zero, we obtain that $t^\star$  satisfies
\begin{align}
\lefteqn{\frac{-\mu e^{-2\mu t^\star}}{N} \sum_{k=0}^{N-1} {}_1 F_1\left(k+1;1;\frac{t^{\star 2}}{N}\right)}&& \nonumber\\
&+ \frac{t^\star e^{-2\mu t^\star}}{N^2} \sum_{k=0}^{N-1}  (k+1) \, _1F_1\left(k+2;2;\frac{t^{\star 2}}{N}\right)=0\ . \nonumber\\\label{trans}
\end{align}
We have solved  Eq. \eqref{trans} numerically for $N<400$ and obtained that   $t^\star$  is well fitted by $t^\star(N)  \approx \alpha N^{\beta}$  with a fitted exponent  $\beta \approx 1/2$, see Fig.~\ref{Fig6}.  A careful asymptotic analysis of Eq. \eqref{trans} is beyond the scope of this paper and is left for future investigation.

\section{Derivation of Eqs. \eqref{StBimodal} and \eqref{StBimodalT} for $S(t)$ in the case of bimodal disorder}\label{appB}
We explicitly solve  the contour integral in Eq. \eqref{finalformula} for the case of a bimodal distribution 
\begin{equation}
p_D(x)=(1-q)\delta(x+\mu_1)+q\delta(x+\mu_2)
\end{equation}
with $\mu_1,\mu_2\geq 0$ and $q\in[0,1]$, and derive the  expression for $S(t)$ given by Eqs. \eqref{StBimodal} and \eqref{StBimodalT}.

Without loss of generality we can consider a shifted distribution 
\begin{equation}
\hat p_D(x)=(1-q)\delta(x)+q\delta(x+\mu)
\end{equation} 
since 
\begin{equation}
S(t;\mu_1,\mu_2)=e^{-2\mu_1 t}S(t;0,\mu_2-\mu_1)\ ,\label{conversion}
\end{equation} 
which follows from the variable transformation  $z\to z-\mu_1$ and $w\to w-\mu_1$ in Eq. \eqref{finalformula}.     

Therefore, we  evaluate the quantity 
\begin{equation}
\hat S(t) = S(t;0,\mu_2-\mu_1)\ .
\end{equation}

Using 
\begin{equation}
\int \mathrm{d}x~\hat p_D(x)\frac{1}{(z-x)(w-x)}=\frac{1-q}{wz}+\frac{q}{(w+\mu)(z+\mu)}\ ,
\end{equation}
in Eq. \eqref{finalformula}, we find that  
\begin{equation}
\hat S(t)=
\frac{1}{(2\pi\mathrm{i})^2}\oiint_\gamma \mathrm{d}z\mathrm{d}w~\frac{e^{t(z+w)}}{\left[\frac{1-q}{wz}+\frac{q}{(w+\mu)(z+\mu)}\right]^{-1}-r^2}\ ,\label{finalformulaApp}
\end{equation} 
where $r^2 = c\overline{J^2}$ is as defined in Eq. \eqref{eq:rDef}.

\subsection{Series expression for $\hat S(t)$}
In order to compute  the contour integral in Eq. \eqref{finalformulaApp}, we express $\hat S(t)$ as an infinite series, which we can integrate term by term.  
To this aim, we use the geometric series 
\begin{equation}
\frac{1}{I^{-1}-r^2}=\frac{I}{1-r^2 I}=\frac{1}{r^2}\sum_{m\geq 0} (r^2 I)^{m+1}\ ,
\end{equation}
in Eq. \eqref{finalformulaApp}, and find
\begin{align}
\nonumber &\hat S(t) =\frac{1}{(2\pi\mathrm{i})^2 }\sum_{m\geq 0}r^{2m}\times\\
&\times\varoiint_\gamma\mathrm{d}z\mathrm{d}w~e^{t(z+w)}
\left(\frac{1-q}{wz}+\frac{q}{(z+\mu)(w+\mu)}\right)^{m+1}\ .
\end{align}
Using the binomial theorem, we obtain the double  sum
\begin{equation}
\hat S(t) = \sum_{m\geq 0}r^{2m}\sum_{n=0}^{m+1}\varphi(n,m;q,-\mu;t)\ , \label{doubleseries}
\end{equation}
where we have denoted
\begin{align}
\nonumber \varphi(n,m;q,-\mu;t) &:={m+1\choose n}q^n (1-q)^{m+1-n}\times\\
&\times\left[\frac{1}{2\pi\mathrm{i}}\oint_\gamma\mathrm{d}z\frac{e^{tz}}{z^{m+1-n}(z+\mu)^n}
\right]^2\ . \label{eq:varphiResult}
\end{align} 

\subsection{Performing the contour integral}
To compute the contour integrals in  $\varphi(n,m;q,-\mu;t)$, we need to evaluate the residues at $z=0$ and $z=-\mu$, and sum them up, 
\begin{align}
\lefteqn{\frac{1}{2\pi\mathrm{i}}\oint_\gamma\mathrm{d}z\frac{e^{tz}}{z^{m+1-n}(z+\mu)^n} } & \nonumber\\
&=  \mathrm{Res}_{z=0}\left[\frac{e^{tz}}{z^{m+1-n}(z+\mu)^n}\right]   \nonumber\\ 
& 
 + 
\mathrm{Res}_{z=-\mu}\left[\frac{e^{tz}}{z^{m+1-n}(z+\mu)^n}\right]\ . 
\end{align}

 The two residues are given by
\begin{align}
\lefteqn{\mathrm{Res}_{z=0}\left[\frac{e^{tz}}{z^{m+1-n}(z+\mu)^n}\right]}& \nonumber \\ 
 &=\frac{(-1)^n t^m }{(m-n)!}\mathrm{U}(n,m+1;-\mu t) \ ,\label{Tric1}
\end{align}
and 
\begin{align}
\lefteqn{\mathrm{Res}_{z=-\mu}\left[\frac{e^{tz}}{z^{m+1-n}(z+\mu)^n}\right] }& \nonumber\\
& =\frac{(-1)^{m+1-n} t^{m} e^{-\mu t}}{(n-1)!}  \mathrm{U}(m-n+1,m+1;-\mu t)\ ,\nonumber\\ \label{Tric2}\ 
\end{align}
respectively, 
where 
\begin{align}
\mathrm{U}(a,b;z) &=   \frac{\Gamma(1-b)}{\Gamma(a+1-b)}  {}_1F_1(a;b;z)
\nonumber\\  
& + \frac{\Gamma(b-1)}{\Gamma(a)} z^{1-b} {}_1F_1(a+1-b;2-b;z)  \label{eq:defx}
\end{align}
 is  the Tricomi hypergeometric function (see Eq.~(13.2.7) in Ref.~\cite{dlmf}), $\Gamma(x) = \int^{\infty}_0 y^{x-1}e^{-y}{\rm d}y$ is the Gamma function, and  $_1F_1(a;b;z)$ is the Kummer confluent hypergeometric function as defined in Eq. \eqref{eq:Kummer}.  
  Note that  although the definition given by Eq. \eqref{eq:defx} does not apply for integer $b$, the Tricomi  hypergeometric function exists also for integer $b$ by continuity of the rhs of Eq. \eqref{eq:defx}.

We illustrate the derivation of Eq. \eqref{Tric1}.
Using the Leibniz formula for the derivative of a product, we obtain 
\begin{align}
\nonumber &\mathrm{Res}_{z=0}\left[\frac{e^{tz}}{z^{m+1-n}(z+\mu)^n}\right] =\\
\nonumber &=\frac{1}{(m-n)!}\lim_{z\to 0}\frac{\mathrm{d}^{m-n}}{\mathrm{d}z^{m-n}}e^{tz}(z+\mu)^{-n}=\\
&=\frac{1}{(m-n)!}\lim_{z\to 0}\sum_{s=0}^{m-n}{m-n\choose s}\frac{t^{m-n-s}e^{tz} (-n)^{\underline{s}}}{(z+\mu)^{n+s}}\ ,
\end{align}
where $(a)^{\underline{s}}$ is the falling factorial as defined in Eq. \eqref{eq:falling}.
Taking the limit $z\rightarrow 0$ in Eq. \eqref{Tric2} and rearranging terms with the help of the Kummer transformation~\cite[Eq. (13.2.40)]{dlmf}
 \begin{equation}
 z^{b-1}\mathrm{U}(a,b;z)=\mathrm{U}(a-b+1,2-b;z)\ ,
 \end{equation}
  we arrive at  
Eq. \eqref{Tric1}.

Equation \eqref{Tric2} is obtained using an analogous reasoning.

Finally, thanks to the identity~\cite[Eq. (13.2.41)]{dlmf} the sum of \eqref{Tric1} and \eqref{Tric2} can be simplified as 
\begin{equation}
\frac{1}{2\pi\mathrm{i}}\oint_\gamma\mathrm{d}z\frac{e^{tz}}{z^{m+1-n}(z+\mu)^n} = t^m \,  _1F_1(n;m+1;-\mu t)/m!\ , \label{eq:integralSol}
\end{equation} 
 in terms of the Kummer confluent hypergeometric function defined in Eq. \eqref{eq:Kummer}.
 
 \subsection{Meaning of the different terms}
 Plugging Eqs. \eqref{eq:integralSol} and \eqref{eq:varphiResult} into the double series Eq. \eqref{doubleseries}, provides us with an explicit expression for $\hat S(t)$.   Unfortunately, this mathematical expression is not meaningful yet.    However, we will see that the $n=0$, $n\in\left\{1,2,
\ldots,m\right\}$, and  $n=m+1$ terms in the double series can be given an appealing interpretation, making sense of the formula. 

      Isolating the $n=0$ and $n=m+1$ terms in \eqref{doubleseries}, we obtain
\begin{align}
\nonumber \hat S(t) &=\sum_{m\geq 0}r^{2m}\left[\varphi(0,m;q,-\mu;t)+\varphi(m+1,m;q,-\mu;t)+\right.\\
&\left.\sum_{n=1}^m \varphi(n,m;q,-\mu;t)\right]\ . \label{eq:Spprime}
\end{align}
Using Eqs. \eqref{conversion}  and \eqref{eq:Spprime}, using  the explicit forms
\begin{align}
\varphi(0,m;q,-\mu;t) &=\frac{(1-q)^{m+1} t^{2 m}}{(m!)^2}\ , \\
\varphi(m+1,m;q,-\mu;t) &=\frac{q^{m+1} t^{2 m} e^{-2 \mu t}}{(m!)^2}\ ,
\end{align}
and using the series expressions  
\begin{align}
\sum_{m\geq 0}r^{2m}\frac{(1-q)^{m+1} t^{2 m}}{(m!)^2} &=(1-q) I_0\left(2 r t \sqrt{1-q} \right)\ , \\
\sum_{m\geq 0}r^{2m}\frac{q^{m+1} t^{2 m} e^{-2 \mu t}}{(m!)^2} &=q e^{-2 \mu  t} I_0\left(2  r t \sqrt{q}\right)\ ,
\end{align}
for the modified Bessel function $I_0$, 
we  obtain the final result given by Eqs. \eqref{StBimodal} and \eqref{StBimodalT}.

\section{Spectra of  random graphs with a prescribed degree distribution and bimodal diagonal disorder}~\label{bimodalSpec}
We analyze the spectral properties of adjacency matrices $A$ of weighted random graphs with a prescribed degree distribution  --- as defined in the introduction of this paper -- for the case where the distribution of diagonal elements is given by Eq. \eqref{eq:bimodal}.  

  If the mean in-degree $c>1$, then the spectrum consists of a continuous part with boundary given by Eq. \eqref{boundary} and with outliers solving Eq. \eqref{outlier}.    Substitution of Eq. \eqref{eq:bimodal} in Eq. \eqref{boundary} leads to the relation 
  \begin{equation}
   \frac{ q }{|\lambda_{\mathrm{b}}+\mu_1|^2}  +  \frac{ 1-q}{|\lambda_{\mathrm{b}}+\mu_2|^2}  = \frac{1}{r^2}  \label{eq:boundary2}
  \end{equation}  
  for the boundary of the continuous part of the spectrum.   Analogously, substituting Eq. \eqref{eq:bimodal} in Eq. \eqref{outlier}  we obtain that outliers solve 
  \begin{equation}
\frac{q}{\lambda_{\rm isol}+\mu_1} +\frac{1-q}{\lambda_{\rm isol}+\mu_2}  = \frac{1}{c\overline{ J }}\ .\label{eq:outlier2}
  \end{equation}   
  
   \begin{figure*}[t]
\centering
\subfigure[$\overline{ J}= 2$,  $r^2 = 10$]
{\includegraphics[width=0.3\textwidth]{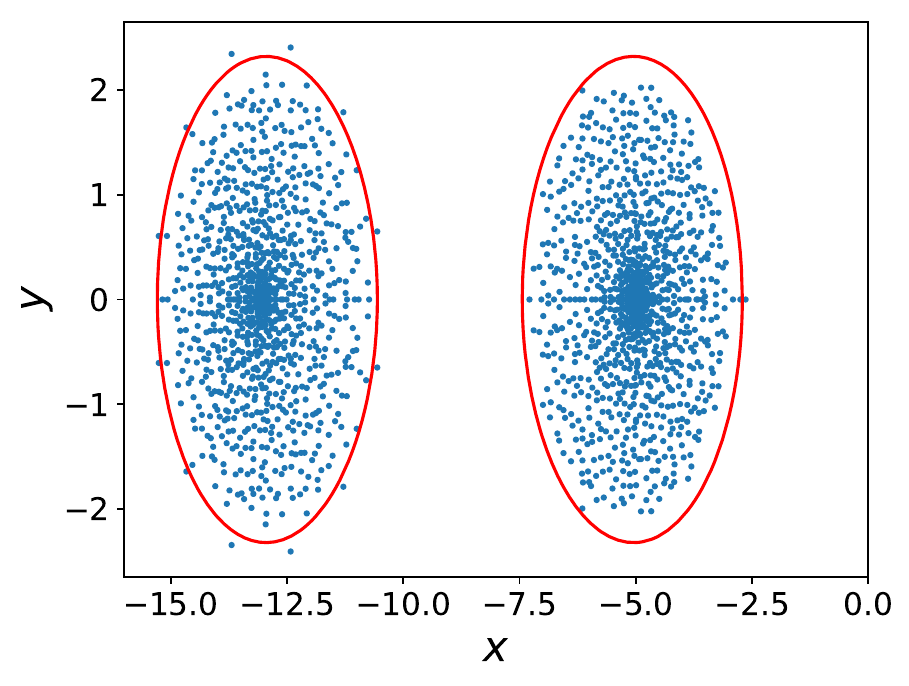}\label{fig:first}} 
\subfigure[$\overline{ J}= 4$,  $r^2 = 34$]
{\includegraphics[width=0.3\textwidth]{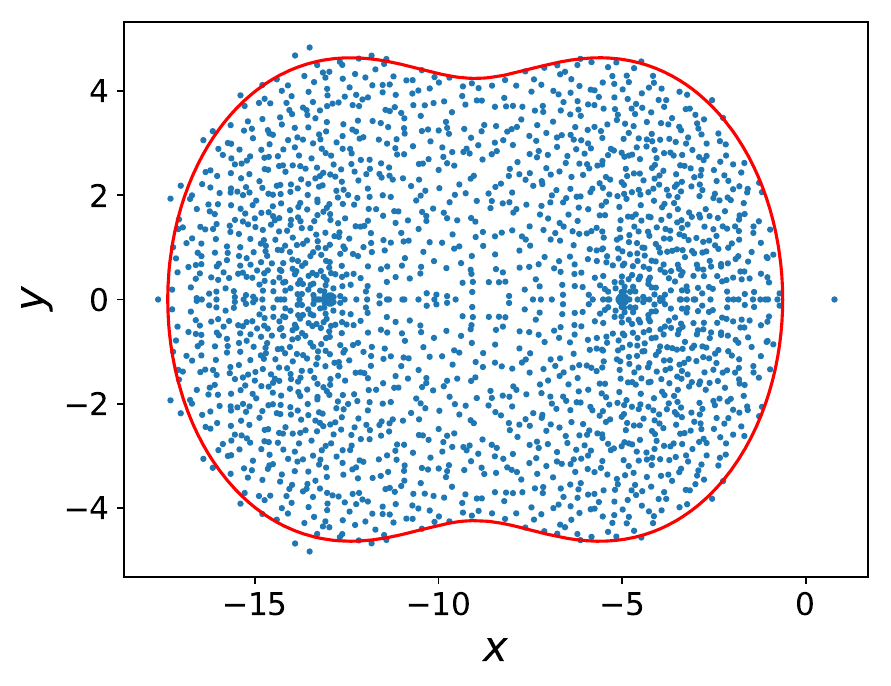}\label{fig:second}} 
\subfigure[$\overline{ J}= 4$, $r^2 = 64$]
{\includegraphics[width=0.3\textwidth]{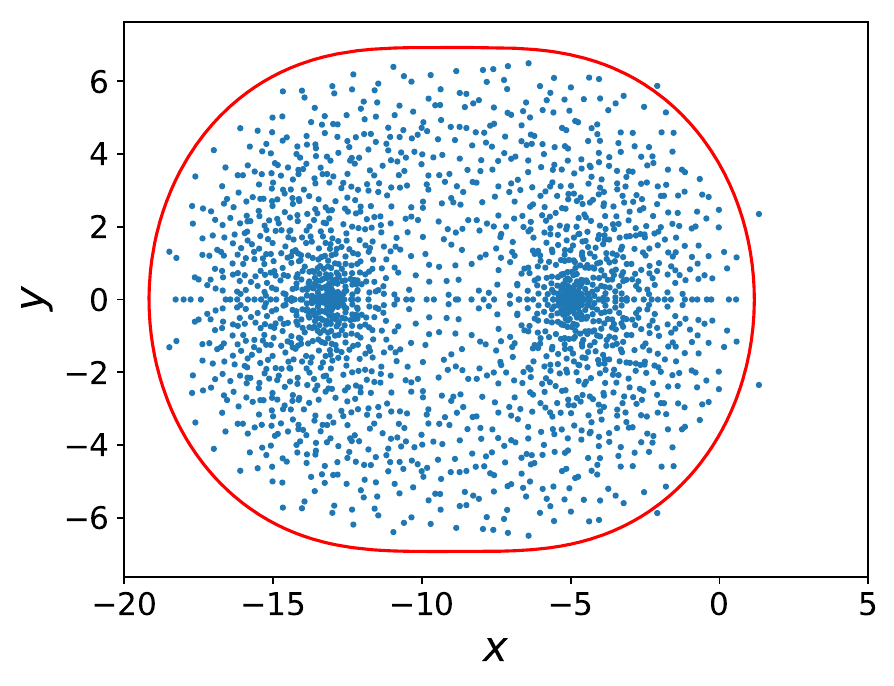}\label{fig:third}} 
\caption{ Spectra of  adjacency matrices of weighted random graphs  with the  bimodal distribution  \eqref{eq:boundary2} on the diagonal, a Gaussian distribution $p_J(x)$, and an in-degree (or out-degree distribution) $p_{\rm deg}(k) = e^{-c}c^k/k!$.   We used the parameters $c=2$, $\mu_1 = 5$, $\mu_2 = 14$, $q = 0.5$ and the values for $\overline{J}$ and $r^2 = c\overline{J^2}$ are provided below the figures.    The red line denotes the theoretical value of the boundary of the absolute continuous spectrum, according to Eq. \eqref{eq:boundary2} while the markers denote the eigenvalues of one single matrix instance of size $N=4000$.         } \label{app:fig3}
\end{figure*}  

  In Fig.~\ref{app:fig3}, we plot the spectra for a few matrix instances  together with the theoretical boundary given by Eq. \eqref{eq:boundary2}.     If $r^2$ is small enough, then  the continuous part of the spectrum consists of two disconnected sets, which are  centered around $-\mu_1$ and $-\mu_2$, respectively.    On the other hand, for large $r$ the continuous spectrum is a simply connected set in the complex plane.  
  
 In order to obtain the leading eigenvalue $\lambda_1$, i.e., the eigenvalue with the largest real part, we determine the real solutions of Eq. \eqref{eq:boundary2}.    If 
 \begin{equation}
 (\mu_1-\mu_2)^2+4r^2 q-4\sqrt{\mathcal{D}}  < 0\ , 
 \end{equation}
   where 
    \begin{equation}
    \mathcal{D} = r^2   \mu_1^2 q - 2 r^2  \mu_1 \mu_2 q +r^2 \mu_2^2 q + r^4  q^2\ , 
  \end{equation}  
  then Eq. \eqref{eq:boundary2} admits two real solutions, namely, 
   \begin{equation}
  \lambda_{\rm b,1} =\frac{-\mu_1-\mu_2 - \sqrt{(\mu_1-\mu_2)^2 + 4r^2 q+4\sqrt{\mathcal{D}}}}{2} 
        \end{equation}
        and
   \begin{equation}
    \lambda_{\rm b,2} =  \frac{-\mu_1-\mu_2 + \sqrt{(\mu_1-\mu_2)^2+4r^2 q +4\sqrt{\mathcal{D}}}}{2}\ .
      \end{equation}
In this scenario, the  continuous part of the spectrum is simply connected.   This is illustrated in panels (b) and (c) of Fig.~\ref{app:fig3}.  On the other hand, if 
       \begin{equation}
 (\mu_1-\mu_2)^2+4r^2 q-4\sqrt{\mathcal{D}}  > 0\ ,
 \end{equation}
 then Eq. \eqref{eq:boundary2} admits four real solutions, namely,  $\lambda_{\rm b,1}$, $\lambda_{\rm b, 2}$, 
   \begin{equation}
    \lambda_{\rm b,3} =  \frac{-\mu_1-\mu_2 -\sqrt{(\mu_1-\mu_2)^2+4r^2 q-4\sqrt{\mathcal{D}}}}{2}
    \end{equation}
    and
    \begin{equation}
        \lambda_{\rm b,4} = \frac{-\mu_1-\mu_2 + \sqrt{(\mu_1-\mu_2)^2+4r^2 q-4\sqrt{\mathcal{D}}}}{2}\ .
  \end{equation}
This the scenario illustrated in panel (a) of Fig.~\ref{app:fig3}. 

  It follows that 
        \begin{equation}
\lambda'_{\rm b} = \max_{\lambda_{\rm b}} \mathrm{Re}[\lambda_{\rm b}]  = \lambda_{\rm b,2}\ .
  \end{equation}    
  
  Solving Eq. \eqref{eq:outlier2}, we find that there may exist two outlier eigenvalues, namely, 
      \begin{align}
\lambda_{1, \rm isol} &=\frac{1}{2} \left(c\overline{J}  -\mu_1-\mu_2 - \mathcal{S}\right)  \\ 
\lambda_{2, \rm isol} &= \frac{1}{2} \left(c\overline{J}  -\mu_1-\mu_2 + \mathcal{S}\right)\ , 
  \end{align}    
  where 
  \begin{equation}
  \mathcal{S} = \sqrt{\left(c\overline{J}  + \mu_1-\mu_2\right)^2   - 4cq\overline{J}  (\mu_1-\mu_2)}\ .
  \end{equation}
  
  Hence, the leading eigenvalue  
  \begin{equation}
  \lambda_1 = {\rm max}\left\{\lambda'_{\rm b} , \lambda_{2, \rm isol} \right\}\ .
  \end{equation}

\section{Power-law random graphs}\label{powerlaw}
Degree distributions of real-world graphs are often  well described by power laws, see for instance Refs.~\cite{Newman2010, Clauset2009, Dorogovtsev2008, Broido, Voitalov}.   Therefore, we compare in this appendix the theoretical result Eq. \eqref{besseldiag} for $S(t)$ with numerical results of $S_N(t)$  on random graphs with a power law degree distribution. 

\begin{figure}[h]
\includegraphics[width=0.45\textwidth]{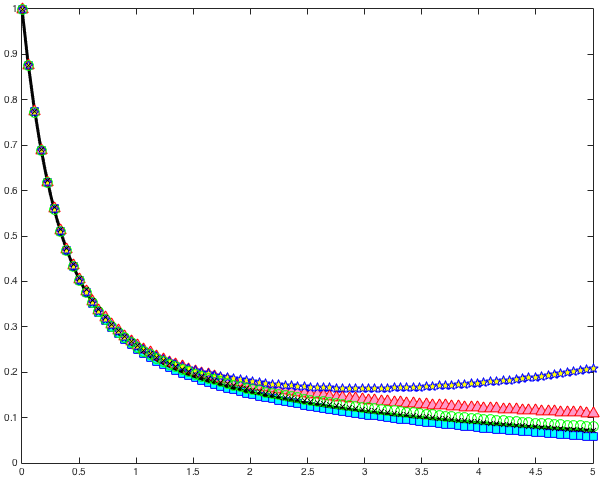}
  \put(-150,-10){$t$}
\caption{ Numerical results for $S_N(t) =\overline{\langle |\bm y(t)|^2\rangle}$ on a power-law random graph   (markers) are compared with the theoretical result   Eq.~\eqref{besseldiag} [given by $S(t) = e^{-2\mu t}I_0(2rt)$] valid for infinitely large graphs (lines).    The prescribed degree distribution is given by Eq. \eqref{eq:powerlaw}, the   couplings $J_{ij}$ are i.i.d.~random variables drawn from the distribution $p_J(x) = (1/2) \delta_{x,1} + (1/2) \delta_{x,-1}$,  $\mu = 1.326$ and $N=2000$.   Markers denote averages of $|\bm{y}(t)|^2$ over five  matrix realizations and 20 initial conditions.  
\label{FigPowerLaw}
}
\end{figure}

In Figure~\ref{FigPowerLaw} we present results for random graphs with the prescribed degree distribution
 \begin{equation}
 p(k_{\rm in},k_{\rm out}) = \frac{k^{-3}_{\rm in} k^{-3}_{\rm out} }{\zeta^2(3)}\ , \label{eq:powerlaw}
 \end{equation} 
 where $\zeta(x)$ is the Riemann-zeta function.   We have set $\overline{J} = 0$  and $\mu$ large enough such  that we are in the stable case, as in sub-panel (a) of Fig.~\ref{Fig1} in the main text   ( all eigenvalues are negative, as ${\rm max_{\lambda_{{\rm b}}}}{\rm Re}[\lambda_{\rm b}] = \sqrt{\zeta(3)/\zeta(2)} - 1.326 < 0$ and there is no outlier).   We observe that the theoretical expression Eq.~\eqref{besseldiag} for $S(t)$ captures very well the empirical results at small enough $t$ and finite $N$.    On the other hand, for larger values of  $t$ we  observe significant sample-to-sample fluctuations.

 We can thus conclude   that the theoretical expression $S(t)$ describes well the dynamical response on random graphs with  $\langle k^\gamma \rangle = \infty$ for $\gamma\geq2$ as long as $t<t^\star(N)$, which is consistent with  results on the leading eigenvalue of directed random graphs with power-law  degree distributions obtained in \cite{Neri2019}.

\end{document}